\documentclass[10pt,journal,compsoc]{IEEEtran}
\ifCLASSOPTIONcompsoc
  \usepackage[nocompress]{cite}
\else
  \usepackage{cite}
\fi
\usepackage{supertabular,booktabs}

\usepackage[T1]{fontenc}
\usepackage{graphicx}
\usepackage[table]{xcolor}
\usepackage{subfigure}
\usepackage{longtable}
\usepackage{multirow}
\usepackage{lscape}
\usepackage{verbatim}
\usepackage{lmodern}
\usepackage[inline]{enumitem}
\usepackage{diagbox}
\usepackage[flushleft]{threeparttable}
\usepackage{tablefootnote}
\usepackage{makecell}
\usepackage{url}
\usepackage{hyperref}

\usepackage{pifont}

\definecolor{jazzberryjam}{rgb}{0.65, 0.04, 0.37}
\definecolor{ultramarine}{rgb}{0.07, 0.04, 0.56}
\definecolor{applegreen}{rgb}{0.55, 0.71, 0.0}
\definecolor{mangotango}{rgb}{1.0, 0.51, 0.26}
\definecolor{mysky}{rgb}{0.1, 0.2, 0.7}
\definecolor{egyptianblue}{rgb}{0.06, 0.2, 0.65}
\definecolor{nicegreen}{rgb}{0.15, 0.76, 0.75}
\definecolor{airforceblue}{rgb}{0.36, 0.54, 0.66}
\definecolor{deepsaffron}{rgb}{1.0, 0.6, 0.2}
\definecolor{electriclime}{rgb}{0.8, 1.0, 0.0}
\definecolor{patriarch}{rgb}{0.5, 0.0, 0.5}
\definecolor{asparagus}{rgb}{0.53, 0.66, 0.42}

\newcommand{\app}[1]{\hyperlink{app:#1}{{\sc App#1}}}
\newcommand{\applabel}[1]{\hypertarget{app:#1}{{\sc App#1}}}

\begin{document}
%%%%%%%%%%%%%%%%%

% \acmConference[x]{y}{z}{t}
% \acmJournal{}
% \acmBooktitle{Under review}
% \editor{}

\title{Fog Computing Applications:\\ Taxonomy and Requirements}

\author{Arif~Ahmed,
  HamidReza~Arkian,
  Davaadorj~Battulga,
  Ali~J.~Fahs,
  Mozhdeh~Farhadi,
  Dimitrios~Giouroukis,
  Adrien~Gougeon,
  Felipe~Oliveira~Gutierrez,
  Guillaume~Pierre,
  Paulo~R.~Souza~Jr,
  Mulugeta~Ayalew~Tamiru,
  Li~Wu,
  \IEEEcompsocitemizethanks{%
    \IEEEcompsocthanksitem A. Ahmed, H. Arkian, A. J. Fahs, A. Gougeon, G. Pierre and P.~R.~Souza~Jr are with Univ Rennes, INRIA, CNRS, IRISA.
    \IEEEcompsocthanksitem  D. Battulga and M. Farhadi are with U-Hopper srl.
    \IEEEcompsocthanksitem  D. Giouroukis and F.O. Gutierrez are with TU Berlin.
    \IEEEcompsocthanksitem  M.A. Tamiru and L. Wu are with Elastisys AB.
  }
  \thanks{All co-authors contributed equally to this paper.}
}

% \author{Arif Ahmed}
% \authornote{All co-authors contributed equally to this paper.}
% \author{HamidReza Arkian}
% \author{Ali Fahs}
% \author{Adrien Gougeon}
% \author{Guillaume Pierre}
% \author{Paulo Souza Jr}
% \affiliation{
%   \institution{Univ Rennes, INRIA, CNRS, IRISA}
%   \city{Rennes, France}
% }
% \author{Davaadorj Battulga}
% \author{Mozhdeh Farhadi}
% \affiliation{
%   \institution{U-Hopper srl}
%   \city{Trento, Italy}
% }
% \author{Dimitrios Giouroukis}
% \author{Felipe Oliveira Gutierrez}
% \affiliation{
%   \institution{Technische Universit\"at Berlin}
%   \city{Berlin, Germany}
% }
% \author{Mulugeta Ayalew Tamiru}
% \author{Li Wu}
% \affiliation{
%   \institution{Elastisys AB}
%   \city{Ume\r{a}, Sweden}
% }
% \renewcommand{\shortauthors}{A. Ahmed \emph{et al.}}
%\keywords{Fog computing, fog applications.}

% \begin{abstract}
%   \notegp{Abstract goes here.}
% \end{abstract}

\maketitle

%\tableofcontents

\begin{abstract}
  Fog computing was designed to support the specific needs of
  latency-critical applications such as augmented reality, and IoT
  applications which produce massive volumes of data that are
  impractical to send to faraway cloud data centers for
  analysis. However this also created new opportunities for a wider
  range of applications which in turn impose their own requirements on
  future fog computing platforms. This article presents a study of a
  representative set of~30 fog computing applications and the
  requirements that a general-purpose fog computing platform should
  support.
\end{abstract}

%%%%%%%%%%%%Introduction%%%%%%%%%%%%%%%%
\section{Introduction}
\label{sec:introduction}

Fog computing extends cloud computing platforms with additional
compute, storage and networking resources that are placed in the
immediate vicinity of end-user devices. Because of its proximity to
end users and their IoT devices, fog computing promises to deliver
extremely low network latencies between the end-user devices and the
fog computing resources serving them, and to process transient data
produced by the end-user devices locally. Research in the domain of
fog computing is currently very active and many researchers propose
new mechanisms to design the next-generation fog computing
platforms~\cite{Yousefpour2019}.

Fog computing researchers however need to face a difficult challenge:
currently, no large-scale general-purpose fog computing platform is
publicly available. To design useful fog computing platforms they,
however, need to understand in detail which kind of applications will
make use of fog computing technologies and which requirements they
will put on the underlying fog platforms. On the other hand, few (if
any) developers will spend significant time building applications
which exploit the capabilities of fog computing platforms unless these
platforms already actually exist.

To break this vicious circle we propose to study a representative set
of fog applications which either were already implemented, or simply
proposed for future development. We carefully selected~30 actual or
proposed applications that cover a wide range of usage types for
future fog computing platforms. We then use this set of reference
applications to address a number of crucial questions about the
functional and non-functional requirements that a general-purpose fog
computing platform should have. We show that fog applications and
their respective requirements are very diverse, and highlight the
specific features that fog platform designers may want to integrate in
their systems to support specific categories of applications.

This article is organized as follows. Section~\ref{sec:background}
presents a general background about fog computing platforms and
applications. Section~\ref{sec:methodology} discusses our methodology
for selecting a representative set of reference applications. Then,
Section~\ref{sec:analysis} analyses the requirements that these
applications put on fog computing platforms and
Section~\ref{sec:conclusion} concludes this article.

%%%%%%%%%%%%%%%%%%%%%%%%%%%%%%%%%%%%%%%%
\section{Background}
\label{sec:background}

Fog computing was originally designed as an extension of cloud
computing with additional compute, storage and communication resources
located close to the end users~\cite{bonomi2012fog,
  vaquero2014finding, openfog-refarch, IEEE-def}. The main purpose of
this technology was to support the specific needs of latency-critical
applications such as augmented reality~\cite{chen2014decentralized},
and IoT applications which produce massive volumes of data that are
impractical to send to faraway cloud data centers for
analysis~\cite{atlam2018fog}. Several excellent surveys of fog
computing technologies are available~\cite{Yousefpour2019, 2017-redowan,
  DBLP:journals/corr/abs-1807-00976}.

However, the expected availability of fog computing technologies has
created opportunities for other types of applications than the
originally anticipated ones. These applications bring their own sets
of requirements which must also be taken into account in the design of
future fog computing platforms.  The published fog computing surveys
focus mostly on fog computing technologies (i.e., the solutions). To
our best knowledge, no article so far has attempted at deriving the
requirements for fog computing platforms from a representative set of
actual and anticipated fog applications. This is the purpose of the
current article.

%%%%%%%%%%%%%%%%%%%%%%%%%%%%%%%%%%%%%%%%
\section{Methodology}
\label{sec:methodology}

We base this study on a review of literature describing specific fog
computing applications. The objective is not to build an exhaustive
list of all proposed applications for fog computing, but rather to
identify a representative sample of typical usages of fog computing
technologies. Because fog computing is still an emerging technology,
we included descriptions of both actual implemented fog application
and proposed future ones. We selected papers based on the following
criteria:
\begin{description}[leftmargin=*,itemsep=0pt,topsep=5pt,labelsep=5pt]
\item[Detailed application description.]  We sought for papers
  containing a detailed technical description of the proposed
  application, and discarded papers which proposed an idea with no
  further technical details.
\item[Publication venue.] We sought for papers which were published in
  peer-reviewed international conferences and journals. In addition,
  we also included white papers published by reputable corporations
  such as Cisco or organizations such as E.U. projects and the OpenFog
  Consortium\footnote{The OpenFog Consortium recently merged with the
    Industrial Internet Consortium. Their use case descriptions are
    temporarily offline while being rebranded to IIC documents. In the
    mean time we make these documents available at
    \url{http://www.fogguru.eu/tmp/OpenFog-Use-Cases.zip}.}.
\item[Economic sectors.] We aimed at identifying applications which
  cover a broad range of economic sectors such as transportation,
  healthcare, entertainment, smart cities, supply chain management,
  smart factories, robotics, agriculture, and security.
\item[No overlap.] To keep the list of application short, we avoided
  including multiple applications which resembled each other too
  much. In such cases we kept the most detailed description in the
  list, and discarded the other similar applications.
\end{description}

\noindent Tables~\ref{tab:refapps1}--\ref{tab:refapps3} show the resulting
list of reference applications.

% \notegp{We need to take a look at this very
% interesting paper~\cite{Gan2019}.}

\begin{table*}[p]
  \centering
  \caption{List of reference applications (1/3)}
  \label{tab:refapps1}
  \begin{tabular}{|l|p{.12\linewidth}|p{.1\linewidth}|p{.1\linewidth}|p{.52\linewidth}|}\hline
  % \hline
    \cellcolor[gray]{0.85}\bf ID & \cellcolor[gray]{0.85}\bf Name & \cellcolor[gray]{0.85}\bf Source & \cellcolor[gray]{0.85}\bf Economic sector & \cellcolor[gray]{0.85}\bf Description \\ \hline
    
     \applabel{01} & LiveMap~\cite{app01} & Peer reviewed &
   Transpor\-tation & \small LiveMap is a scalable mobile information
   system that synthesizes vehicular update streams in real-time. It
   provides fine-grain, deep-zoom details about road conditions and
   hazards such as ``Dead deer in left lane at GPS location (x,y), here
   is an image;'' or, ``Fog detected at GPS location (x,y), visibility
   down to 30 feet, here is a short video clip.'' \\ \hline

   \applabel{02}
   & Wearable cognitive assistance~\cite{app02} & Peer
   reviewed & Health & \small Today, over 20 million Americans are
   affected by some form of cognitive decline. Google glasses integrate
   first-person image capture, sensing, processing and communication
   capabilities. Through context-aware real-time scene interpretation
   (including recognition of objects, faces, activities, signage text,
   and sounds), we can create software that offers helpful guidance for
   everyday life much as a GPS navigation system helps a driver. \\
    \hline

    \applabel{03} & Live video\par broadcasting~\cite{app03} &
  OpenFog Consortium & Entertain\-ment & \small Today's
  sporting events need to broadcast live video from all corners of the
  arena or race course with zero latency. Fans demand to view
  real-time action on their mobile devices over a race course that
  spreads miles over terrain. Hierarchical fog nodes shorten video
  latency and decrease backhaul bandwidth. The Fog delivers the
  agility to manage video services and video algorithms which
  distribute the video process services in different layers from
  camera to cloud, according to their performance requirements.\\
  \hline

  \applabel{04} & Visual security and\par surveillance~\cite{app04}
  & OpenFog Consortium & Smart cities & \small Fog
  computing provides the architecture to build cost-effective,
  real-time and latency-sensitive distributed surveillance systems
  that help to preserve privacy challenges in open environments. Also,
  Fog computing enables real-time tracking, anomaly detection and
  insights from data collected over long time intervals.\\ \hline

  \applabel{05} & Traffic\par congestion\par management~\cite{app05}
  & OpenFog Consortium & Smart cities & \small Fog
  computing gives municipalities a new weapon in the fight against
  traffic congestion.  Fog has the flexibility to leverage
  traffic-related big data, which enables municipalities to take
  measures to alleviate congestion by connecting and analyzing
  previously unconnected infrastructure devices, roadside sensors, and
  on-board vehicles devices, in order to redirect traffic based on
  real-time data.\\ \hline

  \applabel{06} & High-scale drone package delivery~\cite{app06}
  & OpenFog Consortium & Supply chain & \small
  Commercial drones operate in many environments, from aerial to
  subterranean. Fog enables near realtime adjustments and
  collaboration in response to anomalies, operational changes or
  threats. Fog computing enables drones, as self-aware individual fog
  nodes, to interoperate and cooperate as a
  dynamic community.\\ \hline

  \applabel{07} & Process\par Manufac\-turing~\cite{app07} 
  & OpenFog Consortium & Smart\par Factories & \small In order
  to meet market demand, food and beverage producers must be able to
  cope with small-quantity, large-variety products, along with product
  lifecycles with large fluctuations in demand periods and
  quantities. Fog computing helps process brewing by enabling digital
  twins of process in order to replicate key functions, enabling fog
  nodes to scale up or down to meet demand, and ensuring privacy of
  data.\\ \hline

  \applabel{08} & Smart\par Buildings~\cite{app08} &
  OpenFog Consortium & Smart Spaces & \small Today's smart buildings
  are leveraging the IoT for improved business outcomes, such as
  better energy efficiency, improved occupant experience, and lower
  operational costs. They typically contain thousands of sensors
  measuring various building operating parameters such as temperature,
  humidity, occupancy, energy usage, keycard readers and air
  quality. This use case demonstrates how fog nodes at the room level,
  floor level, building level and cloud level can be hierarchically
  architected for efficient real-time processing, enabling dozens of
  new applications. \\ \hline

    \applabel{09} & Real-time Subsurface Imaging~\cite{app09}
  & OpenFog Consortium & Energy-Civil-Environment
  Industry & \small Subsurface imaging and monitoring in real time is
  crucial for understanding subsurface structures and dynamics that
  may pose risks or opportunities for oil, gas and geothermal
  exploration and production. This use case integrates IoT sensor
  networks with fog computing and geophysical imaging
  technology. Fog's scalability enables real-time computation in
  remote field locations, including support for complex compute
  algorithms. \\ \hline

  \applabel{10} & Patient Monitoring~\cite{app10} &
  OpenFog Consortium & Smart Healthcare & \small With its real-time
  communications and analytics requirements for data from thousands of
  low-level sensors, today's hospital patient monitoring requires the
  scalability and agility of fog. This use case is based on a virtual
  compute environment residing on a series of fog nodes that supports
  the flexible deployment of applications and streamlines the
  integration of healthcare systems. \\ \hline
  \end{tabular}
\end{table*}

\begin{table*}[p]
  \centering
  \caption{List of reference applications (2/3)}
  \label{tab:refapps2}
  \begin{tabular}{|l|p{.12\linewidth}|p{.1\linewidth}|p{.1\linewidth}|p{.52\linewidth}|}\hline
    \cellcolor[gray]{0.85}\bf ID & \cellcolor[gray]{0.85}\bf Name & \cellcolor[gray]{0.85}\bf Source & \cellcolor[gray]{0.85}\bf Economic sector & \cellcolor[gray]{0.85}\bf Description \\ \hline

    \applabel{11} & Autonomous Driving~\cite{app11} &
  OpenFog Consortium & Smart Transpor\-tation & \small Autonomous
  Driving, which involves hundreds of simultaneous data processes and
  connections, can't be accomplished without fog. Fog establishes
  trustworthiness of communications between low-level sensors while
  enabling high-bandwidth real-time processing. This use case
  validates how fog architectures for autonomous cars enable
  significantly greater scalability than any other architecture. \\
  \hline

  \applabel{12} & Robots Simultaneous Localization And
  Mapping~\cite{app12} & Peer reviewed &
  Smart\par Robotics & \small By leveraging key principles of fog
  computing that enable processing to take place in close proximity to
  the robots, SLAM is enabled by high-performance real-time edge
  processing, optimized analytics, and heterogeneous applications. The
  SLAM use case speeds up the time to process vast amounts of data
  required in life-or-death situations such as firefighting or rescue
  operations. \\ \hline

  \applabel{13} & Mobility-as-a-Service (MaaS)~\cite{app13} & Cisco
  white paper & Transpor\-tation & \small Fog computing offers the
  potential to understand latent transport demand in real-time, and to
  rapidly assemble insights which can allow MaaS networks to quickly
  deploy services and get people moving. The objective here is to
  deliver a demand-responsive transport ecosystem, where the MaaS
  network enables multiple mobility operators to detect and understand
  customer demand in real-time. \\ \hline

  \applabel{14} & ARQuake~\cite{app14} & Peer reviewed &
  Entertain\-ment & \small ARQuake application is based in the old
  famous shooter called Quake. The augmented reality information
  (monsters, weapons, objects of interest) is displayed in spatial
  context with the physical world using 3D objects. \\ \hline

  \applabel{15} & FAST~\cite{app15} & Peer reviewed & Health & \small
  Stroke (Brain attack) - distributed analytics system to monitor fall
  for stroke mitigation, fall detection algorithms and incorporated
  them into fog-based distributed fall detection system, which
  distribute the analytics throughout the network by splitting the
  detection task between smart phones attached to the users and
  servers in the cloud. \\ \hline

  \applabel{16} & eWall~\cite{app16} & Peer reviewed & Health & \small
  COPD and Mild Dementia are related to aging. eWall provides an
  intelligent home environment with personalized context-aware
  applications based on advanced sensing and fog computing on the
  front and cloud solutions on the back. \\ \hline

  \applabel{17} & Smart street\par lamp~\cite{app17} & Peer reviewed &
  Smart cities & \small Safety and energy consumption of Street Lamp
  is a major concern in Smart
  Cities. % The smart street lamps (SSL) are manually
  % set to either in the state of \emph{switch-on} or \emph{switch-off}.
  % Therefore they have lack of individual control.
  The smart street lamp application deploys various sensors in each
  lamp which collects surrounding real-time data such as the intensity
  of brightness, human presence, voltage level, current level,
  etc. They are also equipped with NB-IoT communication which is used
  to send the collected data to the managing server.  The managing
  server analyzes the collected data and detects any fault in the
  lamp. The computing capacity in the individual lamp enables to
  adjust the brightness intensity of the light depending on the sensor
  data therefore, saves a huge amount of the energy.  \\ \hline

  \applabel{18} & Power\par Consumption Management~\cite{app18} & Peer
  reviewed & Smart Grid & \small Modern home-based IoT devices such as
  electric sensors produce huge amount of data, which are transferred
  to cloud for further processing using long-WAN. The proposed
  implementation offloads some of the cloud tasks to Fog compute nodes
  and therefore reduce latency. The Fog compute nodes monitor the
  usage of electricity for each member to implement the home energy
  management system. \\ \hline

  \applabel{19} & Vehicular Video Processing~\cite{app19} & Peer
  reviewed & Transpor\-tation & \small Processing large volume of
  high-quality videos from vehicles is challenging.  The large volume
  of raw video data is usually sent to the cloud through long-WAN and
  then processed and analyzed and finally, end-users retrieve the
  result from the cloud. The authors study the feasibility of a large
  volume of video transmission at the local fog server which is formed
  combining the computing capacity of collocated vehicles.  \\ \hline

  \applabel{20} & Vehicular Pollution Control~\cite{app20} & Peer
  reviewed & Transpor\-tation & \small This application aims to
  real-time process gas sensor data at the Fog server to reduce the
  latency. It deploys a fog node in each traffic post to process the
  gas sensor data generated from surrounding vehicles. The k-means
  cluster algorithm groups the sensor data and identify the pollution
  level at real-time. If the level of pollution is critical then it
  notified to the Pollution control Board.  \\ \hline 
  \end{tabular}
\end{table*}

\begin{table*}[p]
  \centering
  \caption{List of reference applications (3/3)}
  \label{tab:refapps3}
  \begin{tabular}{|l|p{.12\linewidth}|p{.1\linewidth}|p{.1\linewidth}|p{.52\linewidth}|}\hline
    \cellcolor[gray]{0.85}\bf ID & \cellcolor[gray]{0.85}\bf Name & \cellcolor[gray]{0.85}\bf Source & \cellcolor[gray]{0.85}\bf Economic sector & \cellcolor[gray]{0.85}\bf Description \\ \hline
    
  \applabel{21} & FogLearn~\cite{app21} & Peer reviewed & Healthcare &
  \small FogLearn is a three-layer cloud architecture framework for
  Ganga River Basin Management and for detecting diabetes patients
  suffering from diabetes mellitus. In order to reduce long
  WAN-latency, the Fog layer firstly pre-process the collected data
  from the the Edge Layer and then send to the Cloud for further
  analysis and long term storage.  The Fog Layer also has the capacity
  to scale for analysis the data in case of heavy workload in the
  cloud.  \\ \hline

  \applabel{22} & Telemedecine~\cite{app22} & Peer reviewed & Healthcare & 
  \small To diagnose and evaluate a patient, the healthcare
  professionals need to access the electronic medical record (EMR) of
  the patient, which might contain huge multimedia big data including
  X-rays, ultrasounds, CT scans, and MRI reports. The main focus has
  been given to secure healthcare private data in the cloud using a
  fog computing facility. \\ \hline
  \applabel{23} & SWAMP~\cite{app23} & EU project & Agriculture &
  \small SWAMP develops a high-precision smart irrigation system concept for
  agriculture to enable optimizations of the irrigation system, water
  allocation, and consumption based on a sensor analysis collected
  from the fields. The collected data are transmitted to the central
  fog node deployed in agriculture office through the base station and
  hosted to access the field data to the farmers. However, the
  analysis and modeling of the field data is done in the powerful
  traditional cloud.\\ \hline

  \applabel{24} & Smart Waste Management~\cite{app24} & Peer reviewed
  & Smart cities & \small Waste management is one of the toughest
  challenge that modern cities have to deal with. A city council may
  use sensor data to develop optimised garbage collection strategies,
  so they can save fuel cost related to garbage trucks. \\ \hline

  \applabel{25} & Drone traffic surveillance with
  tracking~\cite{app25} & Peer-reviewed & Security & \small An urban
  speeding traffic monitoring system with tracking using drones, which
  are connect to Fog Nodes in order to process the images. Leveraging
  the divide-andconquer strategy, the subarea containing the vehicle
  of interests was identified and transmitted to the Fog node for
  processing. \\ \hline

  \applabel{26} & GPU-assisted Antivirus Protection in Android
  Devices~\cite{app26} & Peer-reviewed & IT\par security & \small We first
  describe a GPU-based antivirus algorithm for Android devices. Then,
  due to the limited number of GPU-enabled Android devices, we present
  different architecture designs that exploit code offloading for
  running the antivirus on more powerful machines. This approach
  enables lower execution and memory overheads, better performance,
  and improved deployability and management. \\ \hline

  \applabel{27} & Cachier~\cite{app27} & Peer-reviewed &
  Entertain\-ment & \small Recognition and perception based mobile
  applications, such as image recognition, are on the rise. These
  applications are latency-sensitive. Cachier uses the caching model
  along with novel optimizations to minimize latency by adaptively
  balancing load between the edge and the cloud, by leveraging
  spatiotemporal locality of requests, using offline analysis of
  applications, and online estimates of network conditions. \\ \hline

  \applabel{28} & EdgeCourier~\cite{app28} & Peer-reviewed & Broad &
  \small Using cloud storage to automatically back up content changes
  when editing documents is an everyday scenario. EdgeCourier proposes
  the concept of edge-hosed personal service, which has many benefits,
  such as helping deploy EdgeCourier easily in practice. \\ \hline

  \applabel{29} & MMOG~\cite{app29} & Peer-reviewed & Entertain\-ment
  & \small With the increasing popularity of Massively Multiplayer
  Online Game (MMOG) and fast growth of mobile gaming, cloud gaming
  exhibits great promises over the conventional MMOG gaming model as
  it frees players from the requirement of hardware and game
  installation on their local computers.  CloudFog incorporates
  ``fog'' consisting of supernodes that are responsible for rendering
  game videos and streaming them to their nearby players. \\ \hline

  \applabel{30} & Edge Content Caching for Mobile
  Streaming~\cite{app30} & Peer-review & Entertain\-ment & \small
  Increasing popularity of mobile video streaming compels video
  service providers to move from traditional content caching to edge
  network content caching. The authors uses real-world dataset of
  mobile video streaming to study the request pattern and user
  behaviors.  Then analyse the performance of edge content catching in
  the WiFi access network and cellular network. Based on the analysis
  authors proposed an efficient caching strategy in edge
  environment. \\ \hline
  \end{tabular}
\end{table*}

\smallskip The rest of the paper is organized to answer specific
questions about the types of requirements a fog computing system
should fulfill. For each such question, we studied the full list of
applications to build an understanding of the requirements that
different types of applications would have. It is of course up to each
future fog computing systems to decide whether they choose to address
some or all of these requirements.

%%%%%%%%%%%%%%%%%%%%%%%%%%%%%%%%%%%%%%%%
\section{Analysis}
\label{sec:analysis}

\subsection{Why use a fog?}
\label{sec:fog-purposes}

Fog computing was mainly proposed to deliver IoT services (i.e.,
mobility support, context-awareness, geo-distribution and low-latency)
from the edge of the network~\cite{bonomi2012fog}.  By extending cloud
datacenters resources i.e.  compute, storage, and network resources at
the closest vicinity of end users, fog computing also promises to
enhance performance of many applications that require low latency from
IoT devices to their closest fog server, or applications that process
data locally where it is produced~\cite{vaquero2014finding}.

The number of applications running in fog platform is growing. The new
use cases of fog platform are driven by innovative application design
that requires additional platform characteristics which can only be
delivered if the application were deployed next to the end
users~\cite{hong2013mobile}. % We believe that it is necessary to
% identify these performance characteristics so that applications with
% the same requirements can utilize the fog as an underlying platform to
% host the applications.
Therefore, we study all the referred applications and characterize
them based on their reasons for using a fog platform:

% \notepr{Need ref}\noteaa{Added citation}.  We believe that these
% performance characteristics are necessary to identify so that many
% applications with same requirements can utilize fog as an underlying
% platform to host the applications. \notemt{May be better to rephrase
% as: We believe that it is necessary to identify these performance
% characteristics so that applications with the same requirements can
% utilize the fog as an underlying platform to host the applications.}

\begin{description}[leftmargin=*,itemsep=0pt,topsep=5pt,labelsep=5pt]
\item[Reduce latency:] latency-sensitive applications such as
  augmented reality games require end-to-end latency (including
  network and processing delay) under 10-20\,ms~\cite{latency-abrash,
    latency-oculusRift}.  However, the latency between an end user and
  the closest available cloud data center comes in the range of
  20--40\,ms (over wired networks) and up to 150\,ms (over 4G mobile
  networks)~\cite{CLAudit-project}. Therefore, such applications
  cannot realistically run in the cloud.  An obvious solution to
  reduce the end-to-end latency is to deploy the server part of these
  applications in fog platforms.

  % The high
  % latency impact on the run-time performance of such applications
  % for
  % example user's
  % Quality-of-Experience(QoE)~\cite{latency-application}. %\notepr{this
  % sentence
  % seems to be
  % lost here,
  % could you
  % please
  % connect more
  % records. \notemf{Maybe it's better to generalize this
  % sentence. because it is not only a concern for medical records,
  % but also for other types of data... }  %with the
  % text?}
  % \noteaa{Re-written
  % the
  % sentence}.
  % An obvious solution to address this problem is to place cloud
  % server
  % nodes %\notepr{maybe just put server nodes here to not confuse the
  % reader}
  % extremely close to the data within a couple of network hops. For
  % such applications, the server part of the application is deployed
  % in
  % fog platform and therefore reduce the end-to-end latency
  % significantly.

\item[Bandwidth optimization:] edge devices such as IoT sensors, video
  surveillance camera produce large amount of raw data
  everyday~\cite{iot-data, video-surveillance}.  Sending such enormous
  amount of collected data to the cloud for processing creates huge
  network traffic~\cite{shahid:hal-01994156}.  For such applications,
  fog computing plays an intermediate role to reduce the network
  traffic. Fog middlewares are deployed in between the end device and
  the cloud to pre-process the raw collected data at the source and
  only the residual outcome are sent to the cloud for further
  processing~\cite{atlam2018fog}.

\item[Computational offloading:] edge devices such as smart phones,
  smart IoT devices have limited processing capacity.  Running
  compute-intensive applications such as face recognition in those
  devices is painfully slow. Offloading some execution of the
  application to the moderate fog servers may improve the performance.
  Offloading could be another way around: for instance, when a cloud
  server is overloaded, server-side of the running applications cloud
  be offloaded to fog servers~\cite{oueis2015fog}.

%to the fog platform certainly improves the 
%one drawback of edge devices such as
%  smartphones, IoT devices, etc, are hardware limitations. Compute
%  intensive applications, such as face recognition require high
%  processing power. Running compute-intensive application in
%  resource-constrained edge device is painfully slow.
%Although we have seen significant
%  improvement in the latest smartphones hardware \notemf{I think it's better to change the sentence a little bit, in order to include other devices than only smartphones...}. since last decade,
%  they still lack to
%  The ultimate solution is to offload some of the application code to
%  the cloud, execute, and return the result to the edge
%  devices. Therefore, it is possible to solve the incumbency of the
%  weak hardware devices, though the trade-off between user-to-cloud
%  network latency and executing time in the cloud is still fairly
%  discussed ~\cite{kumar2013survey, sharifi2011survey}. However, with
%  the advent of fog computing instead of offloading the application
%  code to the cloud, some of them are executed in the fog platform to
%  gain low-latency performance.  Offloading could be another way
%  around: for instance, when a cloud server is overloaded, the
%  server-side of applications deployed in the cloud can be offloaded
%  to fog servers~\cite{oueis2015fog}.

\item[Privacy and security:] privacy and security are main concerns
  for many applications. For example, E-health applications in health
  care management record enormous amount of patient data for further
  study. Usually, those recorded data are sent to the public cloud for
  the long term storage~\cite{basu2012fusion}.  However, data theft of
  personal medical records is one concerning issue faced by many
  hospitals~\cite{data-theft}. Private fog mitigates the data privacy
  and security issue by delivering storage capacity on-premise of the
  users or the hospitals~\cite{stolfo2012fog}.

\item[Service management:] a growing number of devices like IoT
  sensors and actuators potentially require powerful computing
  machines to operate and manage the devices such as service
  deployment, fault management, hardware installation and device turn
  on/off etc~\cite{biswas2014iot}. Fog computing works as a middleware
  to provide computing power which not only enables one the control of
  the devices but also allows to customize the services based on the
  environment~\cite{taneja2017resource}.

\item[Monitoring edge devices:] monitoring technology has seen
  improvement in hospitals such as monitoring infusion pumps,
  heartbeats, etc.  However, the integration of such devices with
  patients is still challenging which leads to the third-leading cause
  of death each year in the US~\cite{app10}. Fog computing enables to
  remotely host applications that directly communicate with the
  monitoring devices which allow to takes response dynamically based
  on the real-time data~\cite{patient-monitoring}.

\item[Energy efficiency:] energy consumption by large IoT devices
  remains one open issue in the IoT
  environment~\cite{arshad2017green}.  Fog computing enables these
  devices to take decisions intelligently such as switch
  on/off/hibernate that reduces overall energy
  consumption~\cite{jalali2016interconnecting}.

\item[Cost saving:] traditional cloud charges the rented resources
  based on the usage, also known as a pay-you-go model. However, for
  some applications, a one-time investment cost for acquiring the
  private fog resources is preferable to the total cost of cloud
  instances.

\item[Content caching:] content caching or content delivery network or
  content distribution in fog platform is one way to reduce network
  traffic and improve response time by caching popular content
  locally~\cite{buyya2008content}. This practice has been well
  studied, matured, and was able to benefit many
  applications~\cite{pathan2007taxonomy, choi2011survey}.
  Traditionally cloud has been used to deploy content delivery
  network~\cite{papagianni2013cloud}. However, with fog computing, the
  contents can be cached with fine granularity based on the user's
  locality and popularity of the content~\cite{wang2017social}.
\end{description} 

Table~\ref{tab:fog-use-cases} presents a comparison of the surveyed
applications based on the above fog usage. It is not surprising that
most of the applications leverage fog platform to reduce end-to-end
latency and optimizing bandwidth consumption. However one cannot
ignore the other reasons why a fog platform may be used but different
applications.

\begin{table}
  \centering
  \caption{Reasons for using fog computing.}
  \label{tab:fog-use-cases} 
  \begin{tabular}{|p{.23\linewidth}|p{.54\linewidth}|c|}
    \hline
    \cellcolor[gray]{0.85}\bf Fog usages & \cellcolor[gray]{0.85}\bf Applications & \cellcolor[gray]{0.85}\bf Total \\ \hline
    \bf Reduce latency &  \app{02}, \app{03}, \app{04}, \app{06}, \app{08}, \app{09}, \app{11}, \app{12},  \app{13}, \app{14}, \app{15}, \app{16}, \app{17}, \app{18}, \app{20}, \app{21},   \app{25}, \app{27}, \app{29}, \app{30} & 20  \\ \hline
    \bf Bandwidth saving &  \app{01}, \app{02}, \app{04}, \app{05}, \app{07}, \app{08}, \app{11}, \app{14}, \app{15}, \app{16}, \app{18}, \app{19}, \app{20}, \app{21}, \app{23}, \app{24}  & 16 \\ \hline
    \bf Computational offloading &    \app{02},  \app{09}, \app{18}, \app{26}, \app{29}                      &  5 \\  \hline
    \bf Privacy \& security &   \app{10}, \app{18}, \app{22} & 3 \\  \hline
    \bf Service management &   \app{05}, \app{08}, \app{18}, \app{24}, \app{28}    & 5 \\ \hline
    \bf Device monitoring &   \app{09}, \app{10}  & 2 \\ \hline
    \bf Cost saving &  \app{24}   &  1 \\ \hline
    \bf Energy efficiency &  \app{17}    & 1 \\ \hline
    \bf Caching &  \app{27}, \app{30}  &  2\\  \hline
  \end{tabular}
\end{table}

\subsection{Fog deployment models}
\label{sec:fog-models}

We can classify fog models based on the ownership of the fog
infrastructure and underlying resources. There are four
different types of fog models:

\begin{description}[leftmargin=*,itemsep=0pt,topsep=5pt,labelsep=5pt]
\item[Private fog:] A private fog is created, owned, managed and
  operated by some organization, a third party, or some combination of
  them. It may be deployed on or off-premises. The resources of
  private fog are offered for exclusive use by a single organization
  (e.g., business units).

\item[Public fog:] A public fog is created, owned, managed and
  operated by a company, academic institute or government
  organization, or some combination of them.  It is deployed on the
  premises of the fog providers. The resources of public fog are
  offered for open use by the general public.

\item[Community fog:] The community fog is created, managed and
  operated by one or more organizations in the community, also a third
  party, or a combination of them. It may be deployed on or
  off-premises and the resources are offered for exclusive use,
  usually by consumers of a specific community of organizations that
  have shared concerns.

\item[Hybrid fog:] A hybrid fog is a form of fog computing that
  combines the use of public/private/community fog with public/private
  cloud (i.e., hybrid cloud).  It can be useful due to physical
  resource limitations in the fog. Therefore, the platform is extended
  to the hybrid cloud to scale performance. A hybrid cloud is
  scalable, elastic, and resources are available on-demand.
\end{description}

Table~\ref{tab:fog-models} classifies the applications based on the
underlying fog models used to deploy the applications. We found that
nearly half (13 out of 30) of the surveyed applications are deployed
in the \textit{private fog} and the majority (17 out of 30) in the
\textit{hybrid fog}. However, none of the applications are deployed in
\emph{public fog} and \emph{community fog}. Remarkably the
applications can be further categorized based on their requirements
and functionality provided by the respective fog models. We,
therefore, identified the following main reasons for choosing a
\emph{private fog}:

\begin{itemize}[leftmargin=*,itemsep=0pt,topsep=5pt,labelsep=5pt]
\item Privacy and security: many applications that deals with personal
  data such as wearable devices produce a large volume of personal
  data which are too risky to deploy in the open cloud for privacy and
  security concern and, therefore, those applications are usually
  preferred to deploy in a secure fog cloud usually owned by the
  individual or trusted third party.  Similarly, many industries
  prefer to deploy a secure cloud to run automatic robotic application
  due to security reasons.
\item Latency sensitivity and modest resource requirements:
  applications that require moderate resources and low latency are
  deployed in fog platform such as web hosting.
\item Cost-saving: traditional cloud charges rented resources based on
  the usage, also known as a pay-as-you-go model. For some
  applications, the cost is less expensive while deploying the
  application in on-premise, particularly those applications that do
  not need high scalability and maintenance.  For such applications,
  one-time cost for acquiring the private fog resources is cheaper
  than the traditional cloud.
\end{itemize}

A \textit{hybrid fog} mainly aims to scale the resources of fog platforms; 
therefore, applications that require an enormous amount of
resources (i.e., computation, storage, etc.).  We further categorized the
applications deployed in \textit{hybrid fog} based on the following
criteria:

\begin{itemize}[leftmargin=*,itemsep=0pt,topsep=5pt,labelsep=5pt]
\item Compute intensive applications: applications that require
  relatively high computation such as big data analytics in Swarm
  project~\app{23}, face recognition~\app{04} are preferred to deploy
  in hybrid fog.
\item Storage: applications such as~\app{25} that need to store a
  large amount of data for future reference use hybrid fog.
\end{itemize}

\begin{table}
   \centering
   \caption{Fog infrastructure models}
   \label{tab:fog-models} 
   \begin{tabular}{|p{.14\linewidth}|p{.6\linewidth}|c| }
     \hline
     
     \cellcolor[gray]{0.85}\bf Fog model & \cellcolor[gray]{0.85}\bf Applications   								&    \cellcolor[gray]{0.85}\bf Total \\
     \hline		
     \bf Private Fog &  \app{02}, \app{04},  \app{07}, \app{08}, \app{09}, \app{10}, \app{12}, \app{15}, \app{17},  \app{19}, \app{20}, \app{24}, \app{30}  & 13  \\
     \hline		
     
     \bf  Hybrid fog &  \app{01}, \app{03}, \app{05}, \app{06}, \app{11}, \app{13}, \app{14}, \app{16}, \app{18},  \app{21},  \app{22}, \app{23}, \app{25}, \app{26}, \app{27}, \app{28}, \app{29}       & 17  \\
     
     \hline
     
   \end{tabular}
 \end{table}

\subsection{Types of access networks}

Access networks connect the IoT devices to the fog platform, and are
therefore a important basic building blocks for fog platforms. Data
generated from applications need to be processed and acted upon in
terms of milliseconds, therefore, the network architecture should
support ultra-low latency, and large data volumes. There are many
standards and types of access networks that can be deployed on fog
environment. % Physical connectivity, for example, Ethernet is commonly
% used for permanent connections. The data rate using Ethernet can range
% from 10 Mbit/s to 100 Gbit/s depending on the type with an average
% latency of 0.3ms.

Considering the complexity of fog nodes topologies, and due to their
distribution and mobility, wireless connectivity is essential in fog
environment. Wireless connectivity provides flexibility, mobility, and
reachability for levels of hierarchies of fog communication.

As mentioned in the OpenFog Reference Architecture \cite{openfogRA},
wireless support at the fog node will depend on variety of parameters:
function and position in the hierarchy, mobility, coverage, range,
throughput and data transfer rates, etc.

\begin{table*}
  \centering
  \caption{Types of access networks.} 
  \label{tab:accessnetworks}
  \begin{tabular}{ 
    |p{.16\textwidth}
    |p{.24\textwidth}
    |p{.10\textwidth}
    |p{.10\textwidth}
    |p{.15\textwidth}
    |p{.10\textwidth}| 
  }
    \Xhline{2\arrayrulewidth}
    \cellcolor[gray]{0.85}\bf Access Network & \cellcolor[gray]{0.85}\bf Technology & \cellcolor[gray]{0.85}\bf Frequency & \cellcolor[gray]{0.85}\bf Transfer rate & \cellcolor[gray]{0.85}\bf Range & \cellcolor[gray]{0.85}\bf Power consumption \\
    \Xhline{2\arrayrulewidth}
    \multirow{3}{*}{\bf LPWAN \cite{margelis2015low}}
    & LoRaWAN & $\sim$900MHz & 0.3-50kbit/s & 2-5km (urban) 15km (rural) \cite{centenaro2016long} & low \\\cline{2-6}
    & SigFox & 900MHz & 10-1000bit/s & 3-10km (urban) 30-50km (rural) \cite{centenaro2016long} & low \\\cline{2-6}
    & NB-IoT \cite{sinha2017survey} & various & 250 kbit/s & <35km & low \\
    \Xhline{2\arrayrulewidth}
    \multirow{2}{*}{\bf Cellular}
    & 4G LTE \cite{rg4GLTE} & 700-2600MHz & 100Mbit/s & 1-10km & high \\\cline{2-6}
                       & 5G & various~\cite{FGr16band}
                                                        & 700Mbit/s \cite{5Gtest} & less than 4G & high \\
    \Xhline{2\arrayrulewidth}
    \multirow{4}{*}{\bf Wi-Fi}
    & Wi-Fi 4 IEEE 802.11n \cite{abdelrahman2015comparison} & 2.4GHz 5GHz & 150Mbit/s & 70m (indoors) 250m (outdoors) & high \\\cline{2-6}
    & Wi-Fi 5 IEEE 802.11ac \cite{abdelrahman2015comparison} & 5GHz & 860Mbit/s  & 35m (indoors) & high \\\cline{2-6}
    & HaLow IEEE 802.11ah \cite{ahmed2016comparison} & sub 1GHz & 78Mbit/s & 1000m & medium \\\cline{2-6}
    & Wi-Fi 6 IEEE 802.11ax \cite{cisco802.11ax} & 2.4GHz 5Ghz & 600-1800Mbit/s & 76m (indoors) & medium \\
    \Xhline{2\arrayrulewidth}
    \multirow{2}{*}{\bf 802.15.4 based}
    & Zigbee & 2.4GHz & 250kb/s & 10-100m & low \\\cline{2-6}
    & 6LoWPAN & $\sim$900MHz 2.4GHz & 250kb/s & 10-100m & low \\\cline{2-6}
    \Xhline{2\arrayrulewidth}
    \multirow{3}{*} {\bf Other PANs \cite{al2017internet}}
    & Bluetooth Low Energy (BLE) & 2.4GHz & 1Mbit/s & 15-30m & low \\\cline{2-6}
    & RFID & 125kHz 13.56MHz 902-928MHz & 4Mbit/s & <200m & very low \\\cline{2-6}
    & NFC & 125Khz 13.56Mhz 860Mhz  & 106-424kbit/s & 10cm & very low \\
    \Xhline{2\arrayrulewidth}
  \end{tabular}
\end{table*}

Various wireless technologies were classified in
Table~\ref{tab:accessnetworks}, depending on their frequency,
coverage, data transfer rate, and power consumption:
\begin{itemize}[leftmargin=*,itemsep=0pt,topsep=5pt,labelsep=5pt]
\item Low Power Wide Area Network (LPWAN) is a protocol for
  resource-constrained devices and networks over long ranges. It
  covers tens of kilometers, and provides low data rate and power
  usage. Agriculture is the perfect use case for using LPWAN
  technology. Some of the protocols based on LPWAN are LoRa and
  SigFox.
\item Cellular networks are suitable for long distance communications
  in IoT applications. However, all mobile network protocols come at a
  high price due to their licensed Radio Frequency, intellectual
  property protection, and high power consumption.  NB-IoT and LTE-M
  standards are aimed at providing low-power, low-cost IoT
  communication options using existing cellular networks, which can be
  categorized as LPWAN technology. The 5th Generation cellular network
  (5G) is starting to get commercialized, and will improve IoT
  communications. It also promises to lower costs, battery
  consumption, and latency.
\item There exists a wide range of devices with Wi-Fi compatibility,
  with IEEE 802.11 being the most popular network protocol for Local
  Area Networking (LAN). Its high power usage, high data transfer
  rate, and medium-range make it a popular option for latency-aware
  fog applications. HaLow (IEEE 802.11ah) and Wi-Fi 6 (IEEE 802.11ax),
  were designed to address the constraints of IoT networks.
\item The traditional frame format of MAC layer protocols was not
  suitable for IoT low power and multi-hub communications. IEEE
  802.15.4 was created with a more efficient frame format that has
  become the most used IoT MAC layer standard. Applications such as
  home automation are suitable to use low data rate, medium range
  Zigbee, 6LoWPAN technologies.
\item Lastly, other short range Personal Area Network (PAN)
  technologies such as Near Field Communication (NFC), Bluetooth Low
  Energy (BLE), and Radio Frequency Identification (RFID) can be used
  for personal IoT devices like wearable health and fitness trackers,
  asset tracking, check-in systems.
\end{itemize}

Depending on the area that must be covered by the fog platform,
different types of antennas can be proposed. For example, directed
Wi-Fi antennas can be used for Live Video Broadcasting application
\app{04}.

\begin{table}
  \centering
  \caption{Classification of applications based on access network types.} 
  \label{tab:appaccessnetwork}
  \begin{tabular}{ 
    |p{.15\linewidth}
    |p{.63\linewidth}|c|
  }
    \hline
    \cellcolor[gray]{0.85}\bf Access Network & \cellcolor[gray]{0.85}\bf Applications & \cellcolor[gray]{0.85}\bf Total \\
    \hline
    \bf LPWAN & \app{01}, \app{08}, \app{09}, \app{20}, \app{21} & 5 \\
    \hline
    \bf Cellular & \app{01}, \app{03}, \app{05}, \app{06}, \app{09}, \app{11}, \app{12}, \app{13}, \app{15}, \app{23}, \app{19}, \app{25}, \app{26}, \app{29}, \app{30} & 15\\
    \hline
    \bf Wi-Fi & \app{03}, \app{04}, \app{06}, \app{07}, \app{08}, \app{10}, \app{12}, \app{14}, \app{17}, \app{22}, \app{25}, \app{26}, \app{27}, \app{28}, \app{29}, \app{30} & 16\\
    \hline
    \bf 802.15.4 based & \app{07}, \app{08}, \app{12}, \app{16}, \app{17}, \app{18}, \app{22}, \app{24} & 8\\
    \hline
    \bf Other PANs & \app{02}, \app{07}, \app{10}, \app{15}, \app{22}, \app{24} & 6\\
    \hline
  \end{tabular}
\end{table}

Table~\ref{tab:appaccessnetwork} illustrates the possible access
network usages for each application. Some applications are shown
multiple times based on their usage of different access network types
on levels of deployment. Also some of the applications have already
been evaluated in real life testbed. For example, Power Consumption
Management application (\app{22}) proposed to use Zigbee, Vehicle
Video Processing (\app{23}) examined Dedicated Short Range
Communication (DSRC) and LTE, and deployed promising Vehicular Fog
Computing (VFC) architecture.

\subsection{Hardware platforms}
\label{sec:hardware-platform}

The servers which constitute a fog computing platform may be highly
heterogeneous, not only physically but also in terms of resource
capacity such as processing, storage, and network
bandwidth~\cite{marin2017we, adhatarao2017fogg}.  They are considered
the building blocks of fog infrastructures~\cite{iorga2018fog}. Unlike
in traditional cloud platforms, fog architectures are composed of
large number of small computing fog nodes or servers which are placed
in strategic locations across a vast geographical area to cover a
large number of users~\cite{ahmed2018docker}.  Since the fog nodes can
be deployed anywhere between the end users and the cloud, the latency
between an user and the nearest fog nodes largely depends on the
location where the fog nodes are
deployed~\cite{Peterson:2019:DNE:3336937.3336942, fahs2019proximity}.
Depending on the application requirements, application developers have
to choose appropriate fog nodes to improve the QoS of the
applications. We therefore explore different fog nodes used to deploy
the surveyed applications.

Table~\ref{tab:fog-nodes-char} compares different characteristics of
\emph{potential} fog nodes.  We broadly classify the fog nodes into
static and mobile nodes.

\begin{description}[leftmargin=*,itemsep=0pt,topsep=5pt,labelsep=5pt]
\item[Static nodes] are placed in strategic locations. Some examples
  of static nodes are base stations,small-scale datacenters and
  personal laptops, switches, routers, etc. Such devices are
  practically not mobile, therefore need to be placed in a fixed
  location.  Statics nodes can be further categorized based on the
  premise where they are deployed. For example, base stations, network
  resources (switches and routers), small-scale datacenters.

\item[Mobile nodes] are movable, physically small, less resourceful
  and flexible to install and configure.  Some examples of mobile
  nodes are single-board machines~\cite{single-board} (RPIs, Pine
  A64+, etc.), drones, vehicles, etc. They may have small computation
  capacity, however the resources can scale horizontally by
  aggregating nodes~\cite{abrahamsson2013affordable} as discussed in
  Section~\ref{sec:distribution}.
\end{description}

\begin{table*}
  \centering
  \caption{Characteristics of fog nodes.}
  \label{tab:fog-nodes-char} 
  \begin{threeparttable}
    \centering
    \begin{tabular}{|c|c|c|c|c|c|c|c|}
      \hline
      \cellcolor[gray]{0.85}\bf \diagbox[width=13em]{Fog nodes}{Characteristics}& \cellcolor[gray]{0.85}\bf Processing  & \cellcolor[gray]{0.85}\bf Storage & \cellcolor[gray]{0.85}\bf Network & \cellcolor[gray]{0.85}\bf Physical size & \cellcolor[gray]{0.85}\bf Distance from users & \cellcolor[gray]{0.85}\bf Mobility  & \cellcolor[gray]{0.85}\bf Cost \\
      \hline		
      \bf Single-board computers  &   	-	  &		-	&		+/-		&	-	&  -  &    +  &  -   \\
      \hline
      \bf Vehicles (cars, buses..) &   	-	  &		-	&		+/-		&	-	&	-  & +  & -    \\
      \hline
      \bf Drones      			&   	-	  &		-	&		+/-		&	-	&	-  &  +  & +/-   \\
      \hline
      \bf Network resources       &   	-	  &		-	&		+		&	-	&	-  &  +/-    &  +/-  \\
      \hline
      \bf  Laptop / PC 			&   	+/-	  &		+/-	&		+/-		&	+/-	&	+/-  &   -   &  +/-  \\
      \hline
      \bf  Small scale datacenters &   	+	  &		+	&		+		&	+	&	+  &   -   &  + \\
      \hline
    \end{tabular}
    \begin{tablenotes} 
    \item Legends: ``+'' means high, ``+/-'' means neutral, ``-'' means low.
    \end{tablenotes}
  \end{threeparttable}
\end{table*}

Table~\ref{tab:fog-nodes-app} shows the types of fog nodes that are
used to deploy our reference applications. Many of the surveyed
applications may be deployed using one or multiple \emph{potential}
fog nodes depending on various requirements (i.e. computation
capacity, proximity, etc.). We can however draw a number of
conclusions for different types of applications:

\begin{description}[leftmargin=*,itemsep=0pt,topsep=5pt,labelsep=5pt]
\item[Applications focusing on close proximity:] IoT Applications that
  require compute resources at very close proximity of end-users,
  often take the benefits of single-board machines. Some examples of
  such applications are \app{08} and~\app{17} collect sensor data from
  the surrounding environment of the fog nods and process locally.
  Due to the small physical size of single-board machines, they could
  easily be deployed and move from one place to another. Therefore
  applications such as~\app{21} use such small machines to trail
  computation where the robot moves.

\item[Vehicular-based applications:] applications which collect data
  from roadside and process locally such as vehicle video processing,
  autonomous driving, traffic congestion management etc. often take
  advantage of in-built computation capacity of the vehicles.

\item[Drone-based applications:] drone-based applications such
  as~\app{25} and \app{15} take advantage of drones to relocate
  computation resource from one place to another.  The drone is
  equipped with single-board machines that allows to process locally
  and makes communication with other fog nodes.

\item[Compute-intensive applications:] applications such as \app{04}
  and \app{09} require relatively high computation power, therefore
  they are usually deployed in laptops or small-scale datacenters.
\end{description}

\begin{table}
  \centering
  \caption{Fog nodes used to deploy the applications.}
  \label{tab:fog-nodes-app} 
  \begin{tabular}{|p{.2\linewidth}|p{.55\linewidth}|c| }
    \hline
    \cellcolor[gray]{0.85}\bf Fog nodes & \cellcolor[gray]{0.85}\bf Applications  & \cellcolor[gray]{0.85}\bf Total \\
    \hline		
    \bf Single-board computers  &   \app{03}, \app{05},	 \app{07},  \app{08},   \app{11},  \app{12},  \app{17},  \app{18},  \app{21},   \app{23},&	 10  \\
    \hline
    \bf Vehicles (cars, buses..) &    \app{11},	 \app{19},  \app{20}, &	    3 \\
    \hline
    \bf Drones      			&    \app{06},  \app{25},	  &	 2  \\
    \hline
    \bf Network resources       &    \app{03},	  \app{18},  \app{30} &	3  \\
    \hline
    \bf  Laptop / PC 			&   \app{02},  \app{03},  \app{25},  \app{26},  \app{28}	  &	 5  \\
    \hline
    \bf  Small scale datacenters &   \app{01},  \app{04},	 \app{09},  \app{10},  \app{11},   \app{13},   \app{14}, &  16 \\ 
    \bf                         &  \app{15},  \app{16},   \app{17},  \app{18},  \app{22},   \app{23},   \app{24},   \app{27},   \app{29},&	  \\
    \hline
  \end{tabular}
\end{table}

\subsection{Distribution within the fog}
\label{sec:distribution}

Distribution is a key element of fog computing. Locating one node at
the edge of the network is often not sufficient to deliver low
latency, as nodes should be distributed to cover a certain
geographical area. The diffusion of the nodes grants access to nearby
resources for all the users located in a specified area. Fog computing
platforms can provide distribution in two different forms:
\begin{itemize}[leftmargin=*,itemsep=0pt,topsep=5pt,labelsep=5pt]
\item \textbf{Hardware distribution}: portrayed by the distributed
  nodes.
\item \textbf{Software distribution}: portrayed by the distribution of
  the applications' instances and components.
\end{itemize}

For hardware distribution, two common solutions are available. The
first is based on having multiple nodes at the same layer in the
architecture, which is referred to as \textit{Horizontal Distribution}
of the nodes. Meanwhile, if the nodes vary in the size of their
resources (e.g., through some hardware upgrade), then they implement
\textit{Vertical Distribution}. Typically the nodes that have limited
resources are distributed on the edge, however, nodes with more
resources will be placed in a higher vertical layer to serve a greater
number of users. The end of these vertical layers is the cloud with
unlimited resources.

Applications, in general, can be distributed over the cluster either
using \textit{Replication}, where more than one instance of the same
application are placed in different nodes, or by splitting the
application into \textit{Multi-Component}, where each component is
usually a microservice located in a separate node.

Most of the applications intending to run on top of fog try to invest
and benefit from the distribution offered by fog. Some of the
applications that are based on widely-distributed users require a
congruent distribution of the fog nodes. This trend was notable in
IoT-based applications, where for example, in \cite{app17} the IoT
devices cover the lamps in the street, and the nodes should be placed
according to the placement of these devices. This enforces a
horizontal distribution with replication.

We have noticed that some applications rely on software distribution
and more specifically replication of the same components over
different nodes. This approach was evident in the case of
processing-intensive applications like video stream processing. The
reason behind this replication is lowering the latency by placing the
replica in different nodes, and improving performance by providing
more resources~\cite{app19}.

For applications that require vertical distribution, the fog cluster
is layered in a way that the edge node will collect the data, that
will later be sent to the fog nodes which will process the data and
send only the results to the cloud. This architecture is very
efficient in the case of data streams, such that one does not need to
send all the collected data but rather the output results. This is an
effective way to reduce latency and traffic volume over the Internet
connection~\cite{app19}.

Most of the applications that require no distribution at all, are
mainly applications that will run only on the edge or applications
that will load balance between the edge and the cloud.

Table~\ref{tab:Dist} summarizes the surveyed applications and the
types of distribution they require.

\begin{table}[t]
  \centering
  \caption{Hardware and software distribution according to the general category.}
  \label{tab:Dist}
  \begin{tabular}{|c|c|c|c|}
    \hline
    \centering\cellcolor[gray]{0.85}\bf Category & \cellcolor[gray]{0.85}\bf App & \cellcolor[gray]{0.85}\bf Hardware & \cellcolor[gray]{0.85}\bf Software   \\
    \hline
	
    \multirow{3}{*}{ \makecell{\bf Autonomous \\ \bf vehicle}}	& \app{06} & Horizontal & Replication 		\\ \cline{2-4}
    															& \app{11} & Horizontal & Replication 		\\ \cline{2-4}
    															& \app{12} & Vertical 	& Multi-component	\\ \hline
    
    \multirow{2}{*}{ \makecell{\bf Data \\ \bf Storage }} 		& \app{22} & None 		& None  			\\ \cline{2-4}
    															& \app{28} & Vertical 	& None  			\\ \hline
   
    \multirow{14}{*}{ \bf IoT }			 						& \app{07} & Horizontal & Replication 		\\ \cline{2-4}
   																& \app{08} & Both		& Rep. \& Multi. 	\\  \cline{2-4}
    															& \app{09} & Horizontal & Replication 		\\  \cline{2-4}
    															& \app{10} & Horizontal & Replication 		\\ \cline{2-4}
   																& \app{13} & Both		& Replication 		\\  \cline{2-4}
    															& \app{14} & Horizontal & Replication 		\\  \cline{2-4}
    															& \app{15} & Horizontal & Replication 		\\ \cline{2-4}
   																& \app{16} & Horizontal & Replication 		\\  \cline{2-4}
    															& \app{17} & Horizontal & Replication 		\\  \cline{2-4}
    															& \app{18} & Both		& Replication 		\\ \cline{2-4}
   																& \app{20} & Horizontal & Replication 		\\  \cline{2-4}
    															& \app{21} & Horizontal & Replication 		\\  \cline{2-4}
    															& \app{23} & Horizontal & Replication 		\\ \cline{2-4}
   																& \app{24} & Horizontal & Replication 		\\ \hline
    
    \multirow{5}{*}{ \makecell{\bf Real \\ \bf Time }}			& \app{01} & Horizontal & Replication 		\\ \cline{2-4}
   																& \app{02} & None		& None		  		\\  \cline{2-4}
    															& \app{05} & Horizontal & Replication 		\\  \cline{2-4}
    															& \app{25} & Both		& Replication 		\\ \cline{2-4}
   																& \app{26} & None		& None		  		\\ \hline
  
   	\multirow{5}{*}{ \makecell{\bf Media \\ \bf Streaming }}	& \app{03} & Both 		& Rep. \& Multi. 	\\ \cline{2-4}
   																& \app{04} & Horizontal & Replication 		\\  \cline{2-4}
    															& \app{19} & Both 		& Replication 		\\  \cline{2-4}
    															& \app{27} & Vertical	& None 				\\ \cline{2-4}
    															& \app{29} & Horizontal & Replication 		\\ \cline{2-4}
    															& \app{30} & Vertical	& None 				\\ \hline

  \end{tabular}
\end{table}

\subsection{Fog service models}
\label{sec:service-models}

Much like cloud platforms, there are multiple ways by which a fog
computing platform may expose low-level or high-level virtualized
resources to its users. We can classify them in three categories
depending whether they offer infrastructure, platform, and software.
We call them FogIaaS, FogPaaS and FogSaaS to differentiate them from
them cloud-only counterparts.

% These services are available to the end-user and can be accessed at
% real-time through a network. % Depending on the application
% requirements, service providers or application developers can choose
% the appropriate type of fog service before
% renting %\notedg{"consuming" instead of "renting" can be better}
% the resources from fog providers:

%\notedg{"We classify service types based on how they abstract the underlying infrastructure and resources"? this is an alternative, it's up to Arif to decide to keep or remove all the previous sentences and replace with this one.}

\begin{description}[leftmargin=*,itemsep=0pt,topsep=5pt,labelsep=5pt]
\item[FogIaaS (Fog-Infrastructure-as-a-Service)] allows users to take
  advantage of different hardware, such as CPU, network, disk, etc.
  without mentioning the hardware running behind it. The users have
  the independent choice to deploy any Operating System and other
  utilities on the provided resources.

\item[FogPaaS (Fog-Platform-as-a-Service)] provides end-users to
  access basic operating software and optional services for running
  applications and software development environment. FogPaaS builds on
  top of FogIaaS and makes development, testing, and deployment of a
  software quick, robust, and cost-effective.

\item[FogSaaS (Fog-Software-as-a-Service)] provides end-users to use
  software applications without installing them on their personal
  computer.  The services are accessed from the web browser remotely
  through a network.
  % \notedg{"remotely through a network"? I think being too specific with "internet" means that it's the only way to access them, imo}.
\end{description}

Table~\ref{tab:service-models} classifies the reference applications
based on the service models they rely on. In particular, only one
application uses FogSaaS while the others uses either FogPaaS
(10 out of 30) or FogIaaS (19 out of 30).

\begin{table}
  \centering
  \caption{Fog service models.}
  \label{tab:service-models}
  \begin{tabular}{|p{.15\linewidth}|p{.6\linewidth}|c| }
    \hline
    \cellcolor[gray]{0.85}\bf Service models & \cellcolor[gray]{0.85}\bf Applications & \cellcolor[gray]{0.85}\bf Total \\
    \hline
    \bf FogIaaS &   \app{01}, \app{02}, \app{03}, \app{04}, \app{05}, \app{06}, \app{07}, \app{08},  \app{09}, \app{10}, \app{11}, \app{12},  \app{14}, , \app{17}, \app{22}, \app{24}, \app{28}, \app{29} , \app{30} & 19\\ \hline
    \bf FogPaaS &   \app{13}, \app{15},\app{16},  \app{18}, \app{19}, \app{20}, \app{21}, \app{23},   \app{25}, \app{26} & 10 \\ \hline
    \bf FogSaaS &     \app{27}  & 1  \\ \hline
  \end{tabular}
\end{table}

\subsection{Required middlewares}
\label{sec:middleware}

As an increasing number of fog computing applications are being
developed, fog platforms may need to provide greater numbers of
middleware systems (in the form of FogPaaS services) to support easy
application development. We survey the most used types of middlewares
below.

\begin{description}[leftmargin=*,itemsep=0pt,topsep=5pt,labelsep=5pt]
\item[Data stream processing systems] were initially developed by the
  Big Data community~\cite{Carbone2017}.  However, multiple authors
  also recognized their interest in a fog computing environment, where
  they have the potential of reducing data transfers between IoT
  devices and the cloud~\cite{lee2017data}. Multiple systems have been
  developed with a number of variations in their provided
  features~\cite{To:2018:SSM:3296598.3296611}.

\item[Function-as-a-Service] supports the development of event-driven,
  serverless applications.  It enables one to develop, run, and manage
  functions or pieces of code without provision or management of
  servers.  Fog computing is aiming to leverage the possibility to use
  IoT devices in a serverless architecture, which is a type of
  Function as a Service.  This is essentially some extension of the
  Cloud services to the edge where there are the IoT and mobile
  devices, web browsers, and other computing at the edge.

\item[Message-oriented middleware (MOM)] is a method of communication
  between software components in distributed systems.  It can be
  defined as a software or hardware infrastructure which aims to
  support receiving or sending messages between distributed and
  heterogeneous components.  It aims to reduce the complexity of
  developing applications across multiple operating systems and
  network protocols~\cite{curry2004message}.  Message-oriented
  middleware are being used in fog computing environments to improve
  the scalability of Fog nodes and task
  scheduling~\cite{DBLP:journals/corr/GuptaNCG16}.

\item[Web application servers] are software frameworks which provide
  an environment where applications can run regardless of what they
  do.  They usually contain comprehensive service layers where each
  one addresses a separate concern.  Common application servers
  running on the cloud can serve up web pages, provide a container
  model or services for applications, adhere to specification
  controlled by industry, distributed requests across multiple
  physical servers, and provide management and/or development tools.
  Fog computing enhances the capabilities of application servers in
  the direction of facilitating management and programming of
  computing, networking, and storage services between data centers and
  end devices.
\end{description}

Table~\ref{tab:middleware-and-applications} lists the types of
middleware used by the reference applications. Some applications
employ more than one type of middleware, so they are listed multiple
times. Those applications that did not reference any specific type of
middleware are described as unspecified.

\begin{table}[t]
  \caption{Classification of required types of middlewares.}
  \label{tab:middleware-and-applications}
  \begin{tabular}{|p{.2\linewidth}|p{.55\linewidth}|c|}
    \hline
    \cellcolor[gray]{0.85}\bf Required middleware & \cellcolor[gray]{0.85}\bf Applications  & \cellcolor[gray]{0.85}\bf Total \\
    \hline
    \bf Stream Processing Engines& \app{18}, \app{19}, \app{21}, \app{25}, \app{30} & 5 \\
    \hline
    \bf Function as a Service& \app{12}, \app{16}, \app{18}, \app{19}, \app{20}, \app{24}, \app{25}, \app{26}, \app{28}, \app{29} & 10\\
    \hline
    \bf Message-oriented middleware & \app{01}, \app{12}, \app{13}, \app{16}, \app{17}, \app{23}, \app{24} & 7 \\
    \hline
    \bf Web application servers& \app{02}, \app{14}, \app{17}, \app{18}, \app{20}, \app{22}, \app{23}, \app{26}, \app{27}, \app{28}, \app{29} & 11 \\
    \hline
    \bf Unspecified& \app{03}, \app{04}, \app{05}, \app{06}, \app{07}, \app{08}, \app{09}, \app{10}, \app{11}, \app{15} & 20\\
    \hline
  \end{tabular}
\end{table}

\subsection{Types of processed data}
\label{sec:processing}

The type of data processed by an application provides an idea about
the required computing capacity for the nodes constituting the Fog
architecture. This information is closely related to the data volume
produced by the applications and the timeliness of its processing,
detailed in Section~\ref{sec:velocity}.

Fog architectures are often seen as a widely distributed network based
on horizontal scalability~\cite{Yousefpour2019}: if a pool of resource
is insufficient, then a simple solution is to connect more nodes to
the Fog. Nevertheless, the distribution of a single task over several
nodes is rarely mentioned. Nodes are often supposed to be able to host
a full task~\cite{oueis2015fog} and therefore should have adapted
processing and storage capacity to respect the timeliness of
applications.

\begin{table}[t]
  \centering
  \caption{Data processed by applications.}
  \label{tab:dataprocessed}
  \begin{tabular}{|p{.3\linewidth}|p{.45\linewidth}|c|}
    \hline
    \cellcolor[gray]{0.85}\bf Data Type & \cellcolor[gray]{0.85}\bf Applications & \cellcolor[gray]{0.85}\bf Total \\
    \hline
    \bf Textual information only & \app{22} , \app{28} & 2 \\
    \hline
    \bf Sensor information (excluding camera) & \app{07}, \app{08}, \app{10}, \app{13}, \app{15}, \app{16}, \app{17}, \app{18}, \app{20}, \app{21}, \app{23}, \app{24} & 12 \\
    \hline
    \bf Imaging data, CGI or GPU computing (excluding video processing) & \app{09}, \app{26}, \app{27} & 3 \\
    \hline
    \bf Video (and possibly other sensor information) & \app{01}, \app{02}, \app{03}, \app{04}, \app{05}, \app{06}, \app{11}, \app{12}, \app{14}, \app{19}, \app{25}, \app{29}, \app{30} & 13 \\
    \hline
  \end{tabular}
\end{table}

Table~\ref{tab:dataprocessed} classifies the applications according to
their processed data.
% \begin{itemize}
% \item Textual information only;
% \item Sensor information (excluding camera);
% \item Data involving imaging, CGI or other GPU computing (excluding
%   video processing);
% \item Video stream (and possibly other sensor information);
% \end{itemize}
A handful of applications process only textual information, and
arguably do not require very large processing capacity. Applications
processing sensor information may be delay-sensitive. Depending on the
type and number of sensors they may require varying amounts of
processing capacity. We however observed that many of the most
demanding applications actually process either static images or even
video. This often requires the usage of specialized devices such as
GPUs to process incoming (live) video streams in real time.

% From the classification we observe that around half of the
% applications studied process sensor information and more than half of
% them process video streams or require GPU computing.

% Considering those needs on the type of data processed, the nodes in
% multipurpose Fog architectures intending to host several types of
% applications should have important computational capacity and should
% include GPUs.  Nodes with insufficient computing capacity could be
% unable to answer the timeliness of the applications, which is crucial
% in most experiences in a Fog environment.

\subsection{Data volume}
\label{sec:data-volume}

The constant growth of data production is due to the emergence of the
use and collection of sensor data and process automation. Data volume
is the most important aspect in the point of view of the Big Data
community, and it is regarding the amount, size, scale of data
produced and stored~\cite{ISHWARAPPA2015319}.

We observe that few applications use the fog to store significant
amounts of data. Data storage is usually delegated to cloud computing
platforms whereas the fog platforms are mostly used to process streams
of data which are supposed to be processed or filtered quickly,
possibly before being returned to the
users~\cite{adhatarao2017fogg}. % There might be architecture designs
% that the pre-processed data is forwarded to the cloud to be
% appropriately stored.

% Thus, we tried to measure the data volume in fog applications and
% separate them into groups. All the details are collected or inferred
% based on the available pieces of information provided by selected
% applications. The closed groups \emph{What do we mean by \emph{closed
%     groups}} define the amount of required storage per node in a fog
% environment which are reachable into scales of Kilobyte, Megabyte,
% Gigabyte, and Terabyte. \emph{What do we mean by \emph{reachable into
%     scales of ...}} Table~\ref{tab:volume} shows the groups and their
% belonging applications.

\begin{table}[t]
  \centering
  \caption{Data volume.}
  \label{tab:volume}
  \begin{tabular}{|p{.25\linewidth}|p{.5\linewidth}|c|}\hline
    \cellcolor[gray]{0.85}\bf Data Volume Scale & \cellcolor[gray]{0.85}\bf Applications                                                                                                                        & \cellcolor[gray]{0.85}\bf Total \\ \hline
    \bf Kilobytes          & \app{12}, \app{13}, \app{16}, \app{17}, \app{20}, \app{21}, \app{30}                                                                    & 7     \\ \hline
    \bf Megabytes          & \app{02}, \app{05}, \app{06}, \app{10}, \app{14}, \app{15}, \app{19}, \app{24}, \app{26}, \app{27}, \app{28}, \app{29} 					 & 14    \\ \hline
    \bf Gigabytes          & \app{01}, \app{03}, \app{07}, \app{08}, \app{18}, \app{23}, \app{25}, \app{04},                                                                    & 7     \\ \hline
    \bf Terabytes          & \app{09}, \app{11}, \app{22}                                                                                                  & 4     \\ \hline
  \end{tabular}
\end{table}

Table~\ref{tab:volume} shows a classification of the volume of data
processed by each application in broad categories ranging from
kilo-bytes to tera-bytes. We observe that a majority of applications
handle relatively modest amounts of data, in the order of kilo-bytes
to mega-bytes. This is because there is a high demand for real-time
services which relies on receiving, processing, and forwarding the
information without keeping any history or data persistence. For
instance, data stream application or message-oriented applications may
require greater storage capability in situations where it is necessary
to maintain fault tolerance and high parallelism using multiple
replicas. We observe similar strategies for applications which handle
larger volumes of camera-generated data (e.g., \app{04}
and~\app{11}).

Fog applications which belong to the tera-byte scale usually require
persistent and distributed storage in fog nodes (e.g.,~\app{22})
either for keeping history data generated by cameras or personal
files. However, the volume of data they handle largely depends on
their number of users and IoT devices. Applications that use machine
learning or deep learning (e.g., pattern recognition from videos
cameras feeds) may need to frequently retrain their models so storing
large amounts of historical data is important for them. All these
applications usually make use of private for platforms where
sufficient storage capacity may be provisioned for specific
applications.

\subsection{Data velocity and latency sensitivity}
\label{sec:velocity}

One of the fundamental advantages of fog computing is delivering low
user-to-resource latency~\cite{bonomi2012fog,fahs2019proximity}.  The
widely distributed nodes serve this purpose where some of the
resources should be in the vicinity of the end-users. As a result, the
promised low latency of fog is an incentive for applications seeking
ultra-low latencies that may not exceed a couple of milliseconds
regardless of the velocity of the input data.

A very similar notion promoted by the Big Data community is data
velocity, which refers both to the speed at which new input data is
being produced, and to their expected end-to-end processing
latency. In the domain of IoT, the increasing availability of
connected devices and the rapid development of interconnected
applications are leading to continuously increasing rates of input
data that must be processed in a timely fashion~\cite{Gandomi2015}.

% In the domain of Fog Computing, since there are different types of
% applications with different data velocity requirements, an ideal
% infrastructure should be able to handle different data rates coming
% from devices while respecting different latency upper bounds for these
% applications.

\begin{table}[t]
  \centering
  \caption{Data velocity requirements.}
  \label{tab:velocity}
  \begin{tabular}{|p{.22\linewidth}|p{.2\linewidth}|p{.2\linewidth}|p{.2\linewidth}|} \hline
    \cellcolor[gray]{0.85}\bf \diagbox[width=7em]{Latency}{Bandwidth}& \cellcolor[gray]{0.85}\centering\bf kBps or less   & \cellcolor[gray]{0.85}\centering\bf MBps or more   &  \multicolumn{1}{|c|}{\cellcolor[gray]{0.85}\bf Unspecified} \\ \hline
    \bf \centering <10\,ms          & \app{20} \app{06} & \app{03} \app{04} \app{11}  \app{14}  & \app{08} \app{13} \\ \hline
    \bf \centering [10\,ms,100\,ms] & \app{02}  \app{12}  \app{18} & \app{25}  \app{29} \app{30}   & --- \\ \hline
    \bf \centering [100\,ms,1\,s]   & \app{28}          & \app{27}  &  --- \\ \hline
    \bf \centering [1\,s ,10\,s]    & \centering ---    & \app{09}    & --- \\ \hline
    \bf \centering >10\,s           & \app{10}          & \app{01}  \app{23}  & \app{05}  \app{21} \\ \hline
    \bf \centering Unspecified      & \app{07} \app{22} & \centering ---   & \app{15}  \app{16}  \app{17}  \app{19}  \app{24}  \app{26} \\ \hline
  \end{tabular}
\end{table}

We classify the reference applications according to the two dimensions
of data velocity:
\begin{description}[leftmargin=*,itemsep=0pt,topsep=5pt,labelsep=5pt]
\item[Data production speed:] depending on the applications, the input
  data may be produced in the order of kilobytes per second or less,
  megabytes per second, or more.
\item[Expected response latency:] depending on the application, the
  results may be expected within couple of milliseconds, hundreds of
  milliseconds, seconds, or possibly more.
\end{description}

Table~\ref{tab:velocity} classifies the reference applications
according to these two metrics. We can observe that most of the input
data production of the applications are relatively low, in the order
of MBps or even kBps. Such data production rates may initially seem
easy to handle, however, most of the applications require a very quick
response for the generated data. As a result, an increased data
production rate could be challenging for any fog computing platform.

Throughout the reviewed applications, we have noticed certain
trends. While some IoT-based applications such as \app{17} and
\app{20} require no latency restrictions, a significant number of the
applications mention low latency as a vital component for proper
functioning. However, the exact meaning of ``low latency'' varies a
lot depending on the applications. Some applications such as \app{06}
and \app{29} require extremely low latency under 5\,ms, whereas others
such as \app{08} and \app{09} can operate with much softer delivery
time constraints. The applications that require ultra-low latencies
can be classified in two categories:

\begin{description}[leftmargin=*,itemsep=0pt,topsep=5pt,labelsep=5pt]
\item[Real-Time Decision Making:] For these applications such as
  \app{06}, \app{11} and \app{12}, ultra-low latency is a matter of
  human safety. For example, a set of autonomous vehicles calculating
  their trajectory in a fog platform require very fast decisions to
  avoid crashes.
\item[High-Quality User Experience:] Applications falling in this
  category often belong to the entertainment sector for gaming or
  video streaming (\app{02}, \app{29}, \app{30}). In such
  applications, excessive latencies will not cost human lives but may
  seriously affect the users' experience.
\end{description}

Data velocity requirements are an important driver for the design of
any large-scale fog computing platform which must be able to process
input data very close to the location where they have been generated,
and to forward only pre-processed data further to other fog or
cloud-hosted resources.

\subsection{Multiple data providers}
\label{sec:multiactor}

A number of fog applications need to integrate data which originate
from multiple independent data providers. For instance, \app{06}
provides item deliveries through the usage of drone vehicles. This
application may need to know the current location of the recipient as
well as other crucial parts of information, such as weather conditions
or energy-efficient routes. These parameters help the decision-making
process of the application and are usually not available locally
altogether, hence the need to access them from other providers. These
data have different owners and are provided by different companies
with different benefits, possibly with no established trust relation
between them. As external sources are under the control of different
providers, often under different security protocols (e.g., different
keys and algorithms), fog applications may need to deal with different
security mechanisms defined by various data providers. The number of
providers dictates the way the fog application may access and consume
data since the providers employ various data security protocols and
mechanisms, which leads to the application needing to address all
intricacies posed by the different protocols.

Table~\ref{tab:dataactor} classifies our reference applications
according to their data providers. Future fog computing platforms may
need to provide specific mechanisms for the many applications which
rely on multiple independent data providers.

% Since Fog nodes do not have access to huge storage and computation
% resources such as a cloud infrastructure, it is important to specify
% multi-data-provider applications. If an application deals with only
% one data provider, it needs to work with one single security protocol,
% whereas for applications that work with different data providers,
% security management needs more resources as well as mechanisms to
% properly address the different Access Control Lists (ACLs) that
% different providers use and impose to resources. We specify the
% different types of applications in Table~\ref{tab:dataactor}.

\begin{table}
  \centering
  \caption{Data providers.}
  \label{tab:dataactor} 
  \begin{tabular}{|p{.2\linewidth}|p{.52\linewidth}|c|}
    \hline    
    \cellcolor[gray]{0.85}\bf Number of\par Data\par Providers & \cellcolor[gray]{0.85}\bf Applications& \cellcolor[gray]{0.85}\bf Total \\
    \hline    
    \bf Single data\par provider & \app{01}, \app{02}, \app{08}, \app{09}, \app{12}, \app{14}, \app{16}, \app{20} & 8\\
    \hline 
    
    \bf Multiple data providers & \app{03}, \app{04}, \app{05}, \app{06}, \app{07}, \app{10}, \app{11}, \app{13}, \app{15}, \app{17}, \app{18}, \app{19}, \app{20}, \app{22} \app{23}, \app{24} & 16\\
    \hline
  \end{tabular}
\end{table}

\subsection{Privacy sensitivity}
\label{sec:privacy}

In today's societies, data is acting as fuel for most of the services
that we use~\cite{economist}. On the one hand, users are requested to
provide data about themselves in order to receive services. On the
other hand, many citizens are concerned about the usage that may be
made of these data, and prefer not to reveal too much about their
personal life. For example a number of mobile health applications that
receive every day physical activity of their users and provide advice
for healthier life styles. Although the users are interested in such
advices, they do not want their everyday physical activity to be
accessible to their neighbors or colleagues. The European Commission
published the General Data Protection Regulation (GDPR) to address
concerns about the privacy of the citizens~\cite{gdpr}.

Privacy can be defined in many different ways: legal, technical,
societal, etc. In this article we define data privacy as the rules and
acts that can be taken to share personal data to only the legitimate,
intended recipient. Privacy, as a legal term, refers to individuals'
rights to keep their information private and not be accessed by
non-authorized parties.

Because fog applications are located in the immediate vicinity of end
users, they often have access to private user data. Fog computing
therefore inherits all the privacy issues that already existed in IoT
and Cloud computing. Additionally, in Fog computing, we also deal with
data which normally belong to a user, which are in close proximity to
a fog node. For example, a user's location may be inferred from the
location of her closest fog node, which exacerbates the problem of
user privacy.

We categorize data privacy into the following three levels:
\begin{description}[leftmargin=*,itemsep=0pt,topsep=5pt,labelsep=5pt]
\item[Public:] the data may be accessed by anyone. For example, street
  names in a city are public information.
\item[Conditional:] the data may be accessed by some depending on a
  number of conditions. For example, the list of streets in a city
  with higher crime rates may be private or public depending on the
  city's policy.
\item[Private:] the data may be accessed only by a restricted number
  of entities. For example, individual health-related data is usually
  private.
\end{description}

\begin{table}
  \centering
  \caption{Data privacy requirements.}
  \label{tab:privacy} 
  \begin{tabular}{|p{.22\linewidth}|p{.53\linewidth}|c|}
    \hline
    \cellcolor[gray]{0.85}\bf Privacy\par requirements & \cellcolor[gray]{0.85}\bf Applications & \cellcolor[gray]{0.85}\bf Total \\
    \hline    
    \bf Public &   \app{01}, \app{03}, \app{05}, \app{09}, \app{13}, \app{14}, \app{17}, \app{20}, \app{24} & 9 \\
    \hline    
    \bf Conditional & \app{07}, \app{12}, \app{18}, \app{23} & 4
    \\ 
    \hline    
    \bf Private &   \app{02}, \app{04}, \app{06}, \app{08}, \app{10}, \app{11}, \app{15}, \app{16}, \app{19}, \app{21}, \app{22} & 11 \\
    \hline
  \end{tabular}
\end{table}

Table~\ref{tab:privacy} classifies the reference applications with
respect to the privacy level of the data they manipulate.  It is clear
that many applications manipulate private data, which creates
important issues that future fog platforms will have to deal with.

\subsection{Security sensitivity}
\label{sec:security}

An important concern in any large-scale computing infrastructure is
the security of data.  As Fog computing has access to IoT data whether
in the form of data sensed from devices or commands (data) sent from
Cloud to the device, it is important to provide guarantees with
regards to data's integrity and confidentiality. Integrity means that
any sent data should be received intact at its
destination. Confidentiality means that data should only be accessible
to the data source and the legitimate, intended destination.
% Data
% privacy and data confidentiality have similar meaning when we are
% dealing with personal data, but as mentioned in
% section~\ref{sec:privacy}, privacy is considered a legal term. 
As an example for the need of confidentiality, health-related data
sent from an IoT device to a fog application may have high value for
unauthorized organizations, and is therefore at risk for potential
attacks. One such attack happened in 2005 to Anthem where nearly 80
millions users' health records were stolen by an
attacker~\cite{treacherous12}.

Existing solutions utilize encryption as means to provide data
confidentiality. However, cryptography is a compute-intensive process
so not all IoT devices have sufficient processing capacity on
board. In Fog computing we are dealing with heterogeneous IoT devices
with different cryptographic abilities, thus a fog computing platform
may also need to consider the importance of confidentiality as the
other important vector of security in the Fog/IoT space.

In order to avoid data breach, security mechanisms may utilize
context-aware security process where the system alternates between
different security levels according to its own surroundings. As an
example, a node may switch to more powerful encryption schemes because
nodes in proximity to the current one have been marked as infected by
malware.

Table~\ref{tab:security} categorizes the requirements of our reference
applications with respect to data integrity and
confidentiality. Obviously, all applications would prefer operating in
a safe and secure environment. We only list here the applications
where integrity or confidentiality breaches would cause major issues
for their users.

\begin{table}
  \centering
  \caption{Data security requirements.}
  \label{tab:security} 
  \begin{tabular}{|p{.25\linewidth}|p{.5\linewidth}|c|}
    \hline
    \cellcolor[gray]{0.85}\bf Data security\par requirements & \cellcolor[gray]{0.85}\bf Applications& \cellcolor[gray]{0.85}\bf Total \\
    \hline    
    \bf Integrity &   \app{07}, \app{09}, \app{05}, \app{10}, \app{22} & 5\\
    \hline    
    \bf Confidentiality & \app{02}, \app{07}, \app{08}, \app{10}, \app{11}, \app{16}, \app{19}, \app{21}, \app{22}, \app{25}, \app{27}, \app{30} & 12
    \\ 
    \hline
  \end{tabular}
\end{table}

\subsection{Workload characteristics} 

Fog computing platforms are necessarily widely distributed to be
located close to their end users. As such, they experience many
challenges in running highly distributed and large-scale fog
applications as both the fog computing infrastructure itself. To
guarantee the best possible performance and high quality of service,
fog computing platforms need to adapt to dynamic workloads and also be
able to identify and remedy misbehaving workloads.

We can characterize the workload produced by our reference
applications in two general categories and few sub-categories based on
the characteristics of their workloads:

\begin{description}[leftmargin=*,itemsep=0pt,topsep=5pt,labelsep=5pt]
\item[Stable:] the workload is almost static.
\item[Dynamic:] the workload varies according to some criterion:
  \begin{itemize}[leftmargin=*,itemsep=0pt,topsep=5pt,labelsep=5pt]
  \item Location: the workload varies according to the location of fog node.
  \item Time: the workload varies as a function of time.
  \item User: the workload varies according to the load generated by users.
  \end{itemize}
\end{description}

\begin{table}[t]
  \centering
  \caption{Workload characteristics of fog applications.}
  \label{tab:workload}
  \begin{tabular}{|c|p{.13\linewidth}|p{.4\linewidth}|c|}\cline{2-4}
    \Xhline{2\arrayrulewidth}
    \cellcolor[gray]{0.85}\centering\bf Category & \cellcolor[gray]{0.85}\bf Sub\par category & \cellcolor[gray]{0.85}\centering\bf Applications & \cellcolor[gray]{0.85}\bf Total \\
    \Xhline{2\arrayrulewidth}
    \bf Stable  & &  \app{01}, \app{02}, \app{03}, \app{04}, \app{06}, \app{07}, \app{08}, \app{10}, \app{12}, \app{14}, \app{15},  \app{16}, \app{17}, \app{18}, \app{19}, \app{20}, \app{21}, \app{22}, \app{23}, \app{24}, \app{25}, \app{26}, \app{28} & 23 \\ \cline{2-4}
    \Xhline{2\arrayrulewidth}
    \multirow{3}{*}{\rotatebox{0}{ \bf Dynamic }}& Location & \app{05}, \app{09}, \app{11} & 3 \\ \cline{2-4}
    
                                                 & Time & \app{13}  & 1 \\ \cline{2-4}
    
                                                 & User &  \app{27}, \app{29}, \app{30} & 3 \\ 
    \Xhline{2\arrayrulewidth}
  \end{tabular}
\end{table}

Table~\ref{tab:workload} classifies the reference applications
according to their type of workload. We see that the workloads for
most of the fog applications are stable. In these scenarios, sensors
collect data periodically and send them to fog application for
processing. For example, \app{04} uses surveillance cameras to take
pictures at periodic intervals and sends them to the fog for
analysis. These types of applications are present in different sectors
such as transportation, health, entertainment, smart cities, smart
factories, smart buildings, and smart grid.

Seven out of the surveyed applications have dynamic workloads among
which three are dynamic with respect to location, one is dynamic with
respect to time and the rest three are dynamic with respect to
users. Applications that are dynamic with respect to time and user are
the same as typically found in web services running in the
cloud. However, applications that are dynamic with location are
specific to fog environments. When the location changes, a number of
surrounding sensors may change as well as in \app{05}. In \app{09} the
number of self-adaptive added stations may change. In \app{11} the
number of surrounding fog nodes with which the fog node communicates
may changes. Finally, \app{13} collects real-time transport demand in
a mobile way which makes the workload change.

We believe that the awareness of workload characteristics of fog
applications helps in the design of effective and efficient management
and operation of fog computing platforms. Especially for fog
applications with dynamic workloads, it is necessary for the fog
infrastructures and applications to be designed and deployed in a
scalable manner. Meanwhile, fog management platforms need to
incorporate intelligent application placement, dynamic resource
allocation mechanisms and automated operation systems to ensure
acceptable QoS is guaranteed.

\subsection{Implementation maturity}

The reference applications used in this article widely differ in the
level of maturity of their implementation.  We therefore classify them
in four main categories:
\begin{itemize}[leftmargin=*,itemsep=0pt,topsep=5pt,labelsep=5pt]
\item Conceptualization
\item Simulation
\item Prototyping
\item Production
\end{itemize}

\begin{table}[t]
  \centering
  \caption{Maturity level of fog applications.}
  \label{tab:evaluation}
  \begin{tabular}{|c|c|p{.4\linewidth}|c|}\cline{2-4}
    \Xhline{2\arrayrulewidth}
    \cellcolor[gray]{0.85}\centering\bf Maturity & \cellcolor[gray]{0.85}\bf App. & \cellcolor[gray]{0.85}\centering\bf Details & \cellcolor[gray]{0.85}\bf Total \\
    \Xhline{2\arrayrulewidth}
    \multirow{2}{*}{\rotatebox{0}{ \bf Simulation }} &\app{19} & simulated with dataset: Luxembourg SUMO Traffic scenario (LuST) &  \\ \cline{2-3} 
                                    & \app{29} & simulated on Peersim and PlanetLab &  \\  \cline{2-3} 
                                    & \app{30} & simulated with dataset & 3 \\  \cline{2-4}	
    \Xhline{2\arrayrulewidth}
    \multirow{3}{*}{\rotatebox{0}{ \bf Prototyping }}& \app{01} & small scale data center: cloudlet & \\ \cline{2-3} 
    
                                    & \app{02} & cloudlet &  \\ \cline{2-3}   
    
                                    & \app{13} & Cisco Kinetic, with realistic workloads &  \\ \cline{2-3}
    
                                    & \app{14} & wearable computing platform with tracking system &  \\ \cline{2-3}
                                    & \app{17} & self-developed fog platform &  \\ \cline{2-3}
    
                                    & \app{18} & smart socket as the terminal nodes, gateways as the fog node   &  \\ \cline{2-3}
    
                                    & \app{25} & drone as the terminal node,  laptop as the fog node &  \\ \cline{2-3}
                                    & \app{26} & host machine as the edge server &  \\ \cline{2-3}
    
                                    & \app{27} & laptop as the edge server &  \\ \cline{2-3}
    
                                    & \app{28} & laptop as the edge server &  10\\ \cline{2-4}		
    \Xhline{2\arrayrulewidth}
  \end{tabular}
\end{table}

Table~\ref{tab:evaluation} classifies applications according to their
implementation maturity. Interestingly, no application has already
progressed to a ``production'' level of maturity. This shows that the
fog computing technologies and applications still have a long way to
go before they become mainstream. In fact, only 13 out of 30
applications have been developed at all. Three are being simulated and
analyzed with data set or simulation platform, whereas 10 have
actually been prototyped and tested by setting up fog platform or
emulating fog node with similar devices such as cloudlet, laptop, and
gateways. 

% We believe that this analysis of fog applications' implementation
% level could help fog application developers foresee the behaviors of
% fog applications in the real fog platforms. Furthermore, the result
% also provides some useful references to other researchers when
% conducting fog application evaluations.

%%%%%%%%%%%%%%%%%Conclusion%%%%%%%%%%%%%
\section{Conclusion}\label{sec:conclusion}

Fog computing application are extremely diverse, and they logically
impose a very varied types of requirements on the fog computing
platforms designed to support them. Although no current
general-purpose fog platform can pretend addressing all these
requirements, this study aims at helping future platform designers
make informed choices about the features they may or may not support,
and the types of applications that may benefit from them.

\section*{Acknowledgements}

{\small This work is part of the FogGuru project which has received
  funding from the European Union's Horizon 2020 research and
  innovation programme under the Marie Sk\l odowska-Curie grant
  agreement No 765452. The information and views set out in this
  publication are those of the author(s) and do not necessarily
  reflect the official opinion of the European Union. Neither the
  European Union institutions and bodies nor any person acting on
  their behalf may be held responsible for the use which may be made
  of the information contained therein.}

% \bibliographystyle{IEEEtran}
% \bibliography{main}

\begin{thebibliography}{10}
\providecommand{\url}[1]{#1}
\csname url@samestyle\endcsname
\providecommand{\newblock}{\relax}
\providecommand{\bibinfo}[2]{#2}
\providecommand{\BIBentrySTDinterwordspacing}{\spaceskip=0pt\relax}
\providecommand{\BIBentryALTinterwordstretchfactor}{4}
\providecommand{\BIBentryALTinterwordspacing}{\spaceskip=\fontdimen2\font plus
\BIBentryALTinterwordstretchfactor\fontdimen3\font minus
  \fontdimen4\font\relax}
\providecommand{\BIBforeignlanguage}[2]{{%
\expandafter\ifx\csname l@#1\endcsname\relax
\typeout{** WARNING: IEEEtran.bst: No hyphenation pattern has been}%
\typeout{** loaded for the language `#1'. Using the pattern for}%
\typeout{** the default language instead.}%
\else
\language=\csname l@#1\endcsname
\fi
#2}}
\providecommand{\BIBdecl}{\relax}
\BIBdecl

\bibitem{Yousefpour2019}
A.~Yousefpour, C.~Fung, T.~Nguyen, K.~Kadiyala, F.~Jalali, A.~Niakanlahiji,
  J.~Kong, and J.~P. Jue, ``All one needs to know about fog computing and
  related edge computing paradigms: A complete survey,'' \emph{Journal of
  Systems Architecture}, 2019.

\bibitem{bonomi2012fog}
F.~Bonomi, R.~Milito, J.~Zhu, and S.~Addepalli, ``Fog computing and its role in
  the internet of things,'' in \emph{Proc.\ workshop on Mobile computing},
  2012.

\bibitem{vaquero2014finding}
L.~Vaquero, ``Finding your way in the fog: Towards a comprehensive definition
  of fog computing,'' \emph{ACM SIGCOMM Computer Communication Review},
  vol.~44, no.~5, 2014.

\bibitem{openfog-refarch}
{OpenFog Consortium}, ``{OpenFog} reference architecture for fog computing,''
  2017, \url{https://www.openfogconsortium.org/ra/}.

\bibitem{IEEE-def}
{IEEE Standards Association}, ``{IEEE} 1934-2018 -- {IEEE} standard for
  adoption of {OpenFog} reference architecture for fog computing,'' 2018,
  \url{https://standards.ieee.org/standard/1934-2018.html}.

\bibitem{chen2014decentralized}
X.~Chen, ``Decentralized computation offloading game for mobile cloud
  computing,'' \emph{IEEE Transactions on Parallel and Distributed Systems},
  vol.~26, no.~4, 2014.

\bibitem{atlam2018fog}
H.~Atlam, R.~Walters, and G.~Wills, ``Fog computing and the {I}nternet of
  {T}hings: a review,'' \emph{Big Data and Cognitive Computing}, vol.~2, no.~2,
  2018.

\bibitem{2017-redowan}
R.~Mahmud, R.~Kotagiri, and R.~Buyya, \emph{Fog Computing: A Taxonomy, Survey
  and Future Directions}.\hskip 1em plus 0.5em minus 0.4em\relax Springer,
  2018.

\bibitem{DBLP:journals/corr/abs-1807-00976}
\BIBentryALTinterwordspacing
R.~K. Naha, S.~K. Garg, D.~Georgekopolous, P.~P. Jayaraman, L.~Gao, Y.~Xiang,
  and R.~Ranjan, ``Fog computing: Survey of trends, architectures,
  requirements, and research directions,'' \emph{CoRR}, vol. abs/1807.00976,
  2018. [Online]. Available: \url{http://arxiv.org/abs/1807.00976}
\BIBentrySTDinterwordspacing

\bibitem{app01}
W.~Hu, Z.~Feng, Z.~Chen, J.~Harkes, P.~Pillai, and M.~Satyanarayanan, ``Live
  synthesis of vehicle-sourced data over {4G LTE},'' in \emph{Proc.\ ACM
  MSWiM}, 2017.

\bibitem{app02}
K.~Ha, Z.~Chen, W.~Hu, W.~Richter, P.~Pillai, and M.~Satyanarayanan, ``Towards
  wearable cognitive assistance,'' in \emph{Proc.\ MobiSys}, 2014.

\bibitem{app03}
{OpenFog Consortium}, ``Out of the fog: Use case scenarios (live video
  broadcasting),'' 2018, \url{http://www.fogguru.eu/tmp/OpenFog-Use-Cases.zip}.

\bibitem{app04}
------, ``Visual security and surveillance scenario (3.2),'' 2017,
  \url{https://www.iiconsortium.org/pdf/OpenFog_Reference_Architecture_2_09_17.pdf}.

\bibitem{app05}
------, ``Transportation scenario: Smart cars and traffic control (3.1),''
  2017,
  \url{https://www.iiconsortium.org/pdf/OpenFog_Reference_Architecture_2_09_17.pdf}.

\bibitem{app06}
------, ``Out of the fog: Use case scenarios (high-scale drone package
  delivery),'' 2018, \url{http://www.fogguru.eu/tmp/OpenFog-Use-Cases.zip}.

\bibitem{app07}
------, ``Process manufacturing – beverage industry,'' 2018,
  \url{http://www.fogguru.eu/tmp/OpenFog-Use-Cases.zip}.

\bibitem{app08}
------, ``Smart cities scenario (3.3),'' 2017,
  \url{https://www.iiconsortium.org/pdf/OpenFog_Reference_Architecture_2_09_17.pdf}.

\bibitem{app09}
------, ``Real-time subsurface imaging,'' 2018,
  \url{http://www.fogguru.eu/tmp/OpenFog-Use-Cases.zip}.

\bibitem{app10}
------, ``Patient monitoring,'' 2018,
  \url{http://www.fogguru.eu/tmp/OpenFog-Use-Cases.zip}.

\bibitem{app11}
------, ``Autonomous driving,'' 2018,
  \url{http://www.fogguru.eu/tmp/OpenFog-Use-Cases.zip}.

\bibitem{app12}
S.~Dey and A.~Mukherjee, ``Robotic {SLAM}: A review from fog computing and
  mobile edge computing perspective,'' in \emph{Proc. MOBIQUITOUS}, 2016.

\bibitem{app13}
{Cisco}, ``Enabling {MaaS} through a distributed {IoT} data fabric, fog
  computing and network protocols,'' White paper, 2018,
  \url{https://alln-extcloud-storage.cisco.com/ciscoblogs/5c0a6ea91edbb.pdf}.

\bibitem{app14}
B.~Thomas, B.~Close, J.~Donoghue, J.~Squires, P.~D. Bondi, M.~Morris, and
  W.~Piekarski, ``{ARQuake}: An outdoor/indoor augmented reality first person
  application,'' in \emph{Proc.\ International Symposium on Wearable
  Computers}, 2000.

\bibitem{app15}
P.~Hu, S.~Dhelim, H.~Ning, and T.~Qiu, ``Survey on fog computing: architecture,
  key technologies, applications and open issues,'' \emph{Journal of Network
  and Computer Applications}, vol.~98, Nov. 2017.

\bibitem{app16}
S.~Kyriazakos, M.~Mihaylov, B.~Anggorojati, A.~Mihovska, R.~Craciunescu,
  O.~Fratu, and R.~Prasad, ``{eWALL}: an intelligent caring home environment
  offering personalized context-aware applications based on advanced sensing,''
  \emph{Wireless Personal Communications}, vol.~87, no.~3, 2016.

\bibitem{app17}
G.~Jia \emph{et~al.}, ``{SSL}: Smart street lamp based on fog computing for
  smarter cities,'' in \emph{IEEE Transactions on Industrial Informatics},
  vol.~14, no.~11, 2018.

\bibitem{app18}
Y.-D. Chen, M.~Z. Azhari, and J.-S. Leu, ``Design and implementation of a power
  consumption management system for smart home over fog-cloud computing,'' in
  \emph{Proc.\ International Conference on Intelligent Green Building and Smart
  Grid}, 2018.

\bibitem{app19}
C.~Zhu \emph{et~al.}, ``Vehicular fog computing for video crowdsourcing:
  Applications, feasibility, and challenges,'' in \emph{IEEE Communications
  Magazine}, vol.~56, no.~10, Oct. 2018.

\bibitem{app20}
R.~Rajesh and V.~Shijimol, ``Vehicular pollution monitoring and controlling
  using fog computing and clustering algorithm,'' in \emph{International
  Journal of New Innovations in Engineering and Technology}, Mar. 2016.

\bibitem{app21}
R.~K. Barik, R.~Priyadarshini, H.~Dubey, V.~Kumar, and K.~Mankodiya,
  ``{FogLearn}: Leveraging fog-based machine learning for smart system big data
  analytics,'' \emph{International Journal of Fog Computing}, vol.~1, no.~1,
  2018.

\bibitem{app22}
H.~A.~A. Hamid, S.~M.~M. Rahman, M.~S. Hossain, hmad Almogren, and A.~Alamri,
  ``A security model for preserving the privacy of medical big data in a
  healthcare cloud using a fog computing facility with pairing-based
  cryptography,'' \emph{IEEE Access}, vol.~5, 2017.

\bibitem{app23}
{SWAMP Project}, ``Smart water management platform,'' 2018,
  \url{swamp-project.org/communication/WRNP2018_lamina_SWAMP_EN.pdf}.

\bibitem{app24}
C.~Perera, A.~Zaslavsky, P.~Christen, and D.~Georgakopoulos, ``Sensing as a
  service model for smart cities supported by {Internet of Things},''
  \emph{Transactions on Emerging Telecommunications Technologies}, 2013,
  \url{https://doi.org/10.1002/ett.2704}.

\bibitem{app25}
N.~Chen, Y.~Chen, Y.~You, H.~Ling, P.~Liang, and R.~Zimmermann, ``Dynamic urban
  surveillance video stream processing using fog computing,'' in \emph{Proc.\
  BigMM}, 2016.

\bibitem{app26}
D.~Deyannis, R.~Tsirbas, G.~Vasiliadis, R.~Montella, S.~Kosta, and
  S.~Ioannidis, ``Enabling {GPU}-assisted antivirus protection on android
  devices through edge offloading,'' in \emph{Proc.\ EdgeSys}, 2018.

\bibitem{app27}
U.~Drolia, K.~Guo, J.~Tan, R.~Gandhi, and P.~Narasimhan, ``Cachier:
  Edge-caching for recognition applications,'' in \emph{Proc.\ ICDCS}, 2017.

\bibitem{app28}
P.~Hao, Y.~Bai, X.~Zhang, and Y.~Zhang, ``Edgecourier: An edge-hosted personal
  service for low-bandwidth document synchronization in mobile cloud storage
  services,'' in \emph{Proc. ACM/IEEE SEC}, 2017.

\bibitem{app29}
Y.~Lin and H.~Shen, ``{CloudFog}: Leveraging fog to extend cloud gaming for
  thin-client {MMOG} with high quality of service,'' \emph{IEEE Transactions on
  Parallel and Distributed Systems}, vol.~28, no.~2, 2017.

\bibitem{app30}
G.~Ma, Z.~Wang, M.~Zhang, J.~Ye, M.~Chen, and W.~Zhu, ``Understanding
  performance of edge content caching for mobile video streaming,'' \emph{IEEE
  Journal on Selected Areas in Communications}, vol.~35, no.~5, 2017.

\bibitem{hong2013mobile}
K.~Hong, D.~Lillethun, U.~Ramachandran, B.~Ottenw{\"a}lder, and B.~Koldehofe,
  ``Mobile fog: A programming model for large-scale applications on the
  internet of things,'' in \emph{Proc.\ workshop on Mobile cloud computing},
  2013.

\bibitem{latency-abrash}
M.~Abrash, ``Latency – the sine qua non of {AR} and {VR},'' 2012,
  \url{http://blogs.valvesoftware.com/abrash/latency-the-sine-qua-non-of-ar-and-vr/}.

\bibitem{latency-oculusRift}
{OculusRift}, ``Delivers some home truths on latency,''
  \url{https://oculusrift-blog.com/john-carmacks-message-of-latency/682/}.

\bibitem{CLAudit-project}
{CLAudit Project}, ``Planetary-scale cloud latency auditing platform,'' 2016,
  \url{http://bit.do/bS4js}.

\bibitem{iot-data}
{Cisco Inc.}, ``Internet of things (iot) data continues to explode
  exponentially. who is using that data and how?'' 2019,
  \url{https://blogs.cisco.com/datacenter/internet-of-things-iot-data-continues-to-explode-exponentially-who-is-using-that-data-and-how}.

\bibitem{video-surveillance}
{IHS Inc}, ``Big, big, big data: The rise of {HD} video surveillance cameras
  spurs information explosion,'' 2013,
  \url{https://news.ihsmarkit.com/press-release/design-supply-chain-media/big-big-big-data-rise-hd-video-surveillance-cameras-spurs-in}.

\bibitem{shahid:hal-01994156}
M.~R. Shahid, G.~Blanc, Z.~Zhang, and H.~Debar, ``{IoT} devices recognition
  through network traffic analysis,'' in \emph{Proc.\ Big Data}, Dec. 2018.

\bibitem{oueis2015fog}
J.~Oueis, E.~C. Strinati, and S.~Barbarossa, ``The fog balancing: Load
  distribution for small cell cloud computing,'' in \emph{Proc.\ VTC}, 2015.

\bibitem{basu2012fusion}
S.~Basu, A.~H. Karp, J.~Li, J.~Pruyne, J.~Rolia, S.~Singhal, J.~Suermondt, and
  R.~Swaminathan, ``Fusion: managing healthcare records at cloud scale,''
  \emph{Computer}, vol.~45, no.~11, 2012.

\bibitem{data-theft}
{Trend Micro}, ``Healthcare under attack: What happens to stolen medical
  records?'' 2016,
  \url{https://www.trendmicro.com/vinfo/pl/security/news/cyber-attacks/healthcare-under-attack-stolen-medical-records}.

\bibitem{stolfo2012fog}
S.~J. Stolfo, M.~B. Salem, and A.~D. Keromytis, ``Fog computing: Mitigating
  insider data theft attacks in the cloud,'' in \emph{Proc.\ IEEE symposium on
  security and privacy workshops}, 2012.

\bibitem{biswas2014iot}
A.~R. Biswas and R.~Giaffreda, ``{IoT} and cloud convergence: Opportunities and
  challenges,'' in \emph{Proc. IEEE World Forum on Internet of Things}, 2014.

\bibitem{taneja2017resource}
M.~Taneja and A.~Davy, ``Resource aware placement of {IoT} application modules
  in fog-cloud computing paradigm,'' in \emph{Proc.\ Symposium on Integrated
  Network and Service Management}, 2017.

\bibitem{patient-monitoring}
{OpenFog Consortium}, ``Patient monitoring,'' 2019,
  \url{https://www.openfogconsortium.org/wp-content/uploads/Patient-Monitoring-Short.pdf}.

\bibitem{arshad2017green}
R.~Arshad, S.~Zahoor, M.~A. Shah, A.~Wahid, and H.~Yu, ``Green iot: An
  investigation on energy saving practices for 2020 and beyond,'' \emph{IEEE
  Access}, vol.~5, 2017.

\bibitem{jalali2016interconnecting}
F.~Jalali, A.~Vishwanath, J.~De~Hoog, and F.~Suits, ``Interconnecting fog
  computing and microgrids for greening {IoT},'' in \emph{Proc.\ ISGT-Asia},
  2016.

\bibitem{buyya2008content}
R.~Buyya, M.~Pathan, and A.~Vakali, \emph{Content delivery networks}.\hskip 1em
  plus 0.5em minus 0.4em\relax Springer, 2008, vol.~9.

\bibitem{pathan2007taxonomy}
A.-M.~K. Pathan and R.~Buyya, ``A taxonomy and survey of content delivery
  networks,'' \emph{Grid Computing and Distributed Systems Laboratory,
  University of Melbourne, Technical Report}, vol.~4, 2007.

\bibitem{choi2011survey}
J.~Choi, J.~Han, E.~Cho, T.~Kwon, and Y.~Choi, ``A survey on content-oriented
  networking for efficient content delivery,'' \emph{IEEE Communications
  Magazine}, vol.~49, no.~3, 2011.

\bibitem{papagianni2013cloud}
C.~Papagianni, A.~Leivadeas, and S.~Papavassiliou, ``A cloud-oriented content
  delivery network paradigm: Modeling and assessment,'' \emph{IEEE Transactions
  on Dependable and Secure Computing}, vol.~10, no.~5, 2013.

\bibitem{wang2017social}
X.~Wang, S.~Leng, and K.~Yang, ``Social-aware edge caching in fog radio access
  networks,'' \emph{IEEE Access}, vol.~5, 2017.

\bibitem{openfogRA}
{IEEE Standards Association}, ``{IEEE} standard for adoption of {OpenFog}
  reference architecture for fog computing,'' 2018,
  \url{https://standards.ieee.org/standard/1934-2018.html}.

\bibitem{margelis2015low}
G.~Margelis, R.~Piechocki, D.~Kaleshi, and P.~Thomas, ``Low throughput networks
  for the {IoT}: Lessons learned from industrial implementations,'' in
  \emph{Proc.\ WF-IoT}, 2015.

\bibitem{centenaro2016long}
M.~Centenaro, L.~Vangelista, A.~Zanella, and M.~Zorzi, ``Long-range
  communications in unlicensed bands: The rising stars in the {IoT} and smart
  city scenarios,'' \emph{IEEE Wireless Communications}, vol.~23, no.~5, 2016.

\bibitem{sinha2017survey}
R.~S. Sinha, Y.~Wei, and S.-H. Hwang, ``A survey on {LPWA} technology: {LoRa}
  and {NB-IoT},'' \emph{ICT Express}, vol.~3, no.~1, 2017.

\bibitem{rg4GLTE}
G.~S. Nitesh and A.~Kakkar, ``Generations of mobile communication,''
  \emph{International Journal of Advanced Research in Computer Science and
  Software Engineering}, 2016.

\bibitem{FGr16band}
{3GPP}, ``Early progress on rel-16 bands for {5G},'' 2019,
  \url{https://www.3gpp.org/news-events/2025-early-progress-on-rel-16-bands-for-5g}.

\bibitem{5Gtest}
E.~Kauppalehti, ``Real world {4G LTE} vs. {5G} test benchmark: 14x
  bandwidth),'' 2018,
  \url{https://react-etc.net/entry/real-world-4g-lte-vs-5g-test-benchmark-14x-bandwidth}.

\bibitem{abdelrahman2015comparison}
R.~B.~M. Abdelrahman, A.~B.~A. Mustafa, and A.~A. Osman, ``A comparison between
  {IEEE} 802.11 n and ac standards,'' 2015.

\bibitem{ahmed2016comparison}
N.~Ahmed, H.~Rahman, and M.~I. Hussain, ``A comparison of 802.11 ah and 802.15.
  4 for {IoT},'' \emph{ICT Express}, vol.~2, no.~3, 2016.

\bibitem{cisco802.11ax}
{Cisco}, ``{IEEE} 802.11ax the sixth generation of {Wi-Fi},'' 2018,
  \url{https://www.cisco.com/c/dam/en/us/products/collateral/wireless/white-paper-c11-740788.pdf}.

\bibitem{al2017internet}
S.~Al-Sarawi, M.~Anbar, K.~Alieyan, and M.~Alzubaidi, ``Internet of things
  (iot) communication protocols,'' in \emph{Proc.\ ICIT}, 2017.

\bibitem{marin2017we}
E.~Mar{\'\i}n-Tordera, X.~Masip-Bruin, J.~Garc{\'\i}a-Almi{\~n}ana, A.~Jukan,
  G.-J. Ren, and J.~Zhu, ``Do we all really know what a fog node is? current
  trends towards an open definition,'' \emph{Computer Communications}, vol.
  109, 2017.

\bibitem{adhatarao2017fogg}
S.~S. Adhatarao, M.~Arumaithurai, and X.~Fu, ``{FOGG}: A fog computing based
  gateway to integrate sensor networks to internet,'' in \emph{Proc.\ ITC},
  vol.~2, 2017.

\bibitem{iorga2018fog}
M.~Iorga, L.~Feldman, R.~Barton, M.~Martin, N.~Goren, and C.~Mahmoudi, ``Fog
  computing conceptual model, recommendations of the national institute of
  standards and technology,'' \emph{NIST Special Publication}, 2018.

\bibitem{ahmed2018docker}
A.~Ahmed and G.~Pierre, ``Docker container deployment in fog computing
  infrastructures,'' in \emph{Proc.\ IEEE EDGE}, 2018.

\bibitem{Peterson:2019:DNE:3336937.3336942}
L.~Peterson, T.~Anderson, S.~Katti, N.~McKeown, G.~Parulkar, J.~Rexford,
  M.~Satyanarayanan, O.~Sunay, and A.~Vahdat, ``Democratizing the network
  edge,'' \emph{SIGCOMM Comput. Commun. Rev.}, vol.~49, no.~2, May 2019.

\bibitem{fahs2019proximity}
A.~Fahs and G.~Pierre, ``Proximity-aware traffic routing in distributed fog
  computing platforms,'' in \emph{Proc.\ IEEE/ACM CCGrid}, 2019.

\bibitem{single-board}
{Wikipedia}, ``Single board computer,'' 2019,
  \url{https://en.wikipedia.org/wiki/Single-board_computer}.

\bibitem{abrahamsson2013affordable}
P.~Abrahamsson, S.~Helmer, N.~Phaphoom, L.~Nicolodi, N.~Preda, L.~Miori,
  M.~Angriman, J.~Rikkil{\"a}, X.~Wang, K.~Hamily \emph{et~al.}, ``Affordable
  and energy-efficient cloud computing clusters: The {Bolzano Raspberry Pi}
  cloud cluster experiment,'' in \emph{Proc. Intl.\ Conference on Cloud
  Computing Technology and Science}, 2013.

\bibitem{Carbone2017}
P.~Carbone, G.~E. G{\'e}vay, G.~Hermann, A.~Katsifodimos, J.~Soto, V.~Markl,
  and S.~Haridi, \emph{Large-Scale Data Stream Processing Systems}.\hskip 1em
  plus 0.5em minus 0.4em\relax Springer, 2017.

\bibitem{lee2017data}
H.~Lee, J.~Oh, K.~Kim, and H.~Yeon, ``A data streaming performance evaluation
  using resource constrained edge device,'' in \emph{Proc.\ ICTC}, 2017.

\bibitem{To:2018:SSM:3296598.3296611}
Q.-C. To, J.~Soto, and V.~Markl, ``A survey of state management in big data
  processing systems,'' \emph{The VLDB Journal}, vol.~27, no.~6, Dec. 2018.

\bibitem{curry2004message}
E.~Curry, ``Message-oriented middleware,'' \emph{Middleware for
  communications}, 2004.

\bibitem{DBLP:journals/corr/GuptaNCG16}
\BIBentryALTinterwordspacing
H.~Gupta, S.~B. Nath, S.~Chakraborty, and S.~K. Ghosh, ``{SDFog}: a software
  defined computing architecture for {QoS} aware service orchestration over
  edge devices,'' \emph{CoRR}, vol. abs/1609.01190, 2016. [Online]. Available:
  \url{http://arxiv.org/abs/1609.01190}
\BIBentrySTDinterwordspacing

\bibitem{ISHWARAPPA2015319}
{Ishwarappa} and J.~Anuradha, ``A brief introduction on big data {5Vs}
  characteristics and {H}adoop technology,'' \emph{Procedia Computer Science},
  vol.~48, 2015.

\bibitem{Gandomi2015}
A.~Gandomi and M.~Haider, ``Beyond the hype: Big data concepts, methods, and
  analytics,'' \emph{International Journal of Information Management}, vol.~35,
  no.~2, 2015.

\bibitem{economist}
Economist, ``Data is giving rise to a new economy,'' 2017,
  \url{https://www.economist.com/briefing/2017/05/06/data-is-giving-rise-to-a-new-economy}.

\bibitem{gdpr}
{Council of European Union}, ``Council regulation ({EU}) no 269/2014,'' 2014,
  \newline\url{https://eur-lex.europa.eu/legal-content/EN/TXT/HTML/?uri=OJ:L:2016:119:FULL&from=EL}.

\bibitem{treacherous12}
{Cloud Security Alliance}, ``The treacherous 12 -- top threats to cloud
  computing + industry insights,'' 2017,
  \url{https://downloads.cloudsecurityalliance.org/assets/research/top-threats/treacherous-12-top-threats.pdf}.

\end{thebibliography}


\end{document}